\documentclass[alpha-refs]{wiley-article}

\usepackage{color}
\usepackage{ulem}
\usepackage{wrapfig}

\usepackage{siunitx}
\usepackage{algorithm}
\usepackage{algpseudocode}
\usepackage{amssymb}
\usepackage[T1]{fontenc}
\usepackage{amsmath}
\usepackage{mathtools}

\newcommand{\N}{\mathbb{N}}

\newcommand{\R}{\mathbb{R}}

\papertype{Article}

\title{Arnoldi Singular Vector perturbations for machine learning weather prediction}

\author[1,2]{Jens Winkler}
\author[1]{Michael Denhard}

\affil[1]{Deutscher Wetterdienst, Offenbach am Main, Germany}
\affil[2]{Department of Mathematics and Computer Science, Philipps Universität Marburg, Germany}

\corraddress{Jens Winkler, Department for meteorological analysis and modelling, Deutscher Wetterdienst, Frankfurter Straße 135, 63067 Offenbach am Main, Germany}
\corremail{winklerj@mathematik.uni-marburg.de , jens.winkler@dwd.de}

\fundinginfo{
BMV Forschungsnetzwerk, Topic area 3, Reliable transport infrastructure, project "Ensemble forecasts of extreme weather", German Federal Ministry for Transport;\\

European Center for Medium-range Weather Forecasts (ECMWF), Earth System Modelling (ESM) group, staff secondment
}

\runningauthor{Winkler and Denhard}

\begin{document}

\maketitle

\begin{abstract}
Since weather forecasts are fundamentally uncertain, reliable decision making requires information on the likelihoods of future weather scenarios. We explore the sensitivity of machine learning weather prediction (MLWP) using the 24h Pangu Weather ML model of Huawei to errors in the initial conditions with a specific kind of Singular Vector (SV) perturbations. Our Arnoldi-SV (A-SV) method does not need linear nor adjoint model versions and is applicable to numerical weather prediction (NWP) as well as MLWP. It observes error growth within a given optimization time window by iteratively applying a forecast model to perturbed model states. This creates a Krylov subspace, implicitly based on a matrix operator, which approximates the local error growth. Each iteration adds new dimensions to the Krylov space and its leading right SVs are expected to turn into directions of growing errors. We show that A-SV indeed finds dynamically meaningful perturbation patterns for the 24h Pangu Weather model, which grow right from the beginning of the forecast rollout. These perturbations describe local unstable modes and could be a basis to initialize MLWP ensembles. Since we start A-SV from random noise perturbations, the algorithm transforms noise into perturbations conditioned on a given reference state - a process that is akin to the denoising process of the generic diffusion based ML model of GenCast, therefor we briefly discuss similarities and differences.

\end{abstract}

\section{Introduction}

The equations of the atmospheric circulation and the parametrization of the sub-grid scale physical processes preserve the knowledge of generations of scientists. They built the numerical weather prediction (NWP) models, which are the backbone of operational weather forecasting at the national meteorological services worldwide. But recently, data-driven forecasting with artificial neural networks (NN) became competitive \citep{benbouallegue2023risedatadrivenweatherforecasting}. Machine learning weather prediction (MLWP) models learn from long historic datasets of atmospheric states, which are usually taken from the NWP based ERA5 re-analysis dataset of ECMWF \citep{hersbach2020era5}. They can produce realistic atmospheric forecasts, which stay active for quite a long time and partly even show greater forecast skill compared to the best available NWP models \citep{keisler2022forecasting, kurth2022fourcastnet, lam2023learning, chen2023fengwu,nguyen2023scaling, li2023fuxi}. 

However, it has been found that advantages of MLWP upon NWP oftentimes increase with the forecast lead time (see e.g. \cite{benbouallegue2023risedatadrivenweatherforecasting}). This indicates that at least some part of their success is due to the well known ``regression to the mean'' effect of statistical models. When the skill of a forecast gradually vanishes the smoothing of the predicted patterns eliminates unpredictable variability. Otherwise, a fully developed cyclone or squall line in the wrong position put a double penalty on the forecast. NWP accepts double penalties by purpose, because the numerical integrations of the governing equations as well as the physical parametrizations are designed to preserve balances and avoid drifts towards the climate to keep the forecast active. Hence, the commonly preferred strategy for predicting forecast uncertainty in NWP is running ensembles of forecasts with slightly different initial conditions and perturbations of the model physics \citep{kalnay2003atmospheric, palmer2019ecmwf, roberts2023improver, ifs-manual-cy46r1-ens, yamaguchi2018introduction,Zhu2012b,palmer2006predictability}.

In this paper we explore the effects of initial condition uncertainty on the forecast. This uncertainty arises from the imperfectness of the data assimilation process including measurement errors or the fragmentary nature of the observations. Unfortunately, the complexity of the atmospheric circulation prevents a comprehensive sampling of analysis uncertainty, because the number of perturbation experiments with todays NWP models is very limited. Even the much cheaper MLWP inference does not allow to overcome the ``curse of dimensionality''.

The most simple and universal perturbation strategy is adding small amounts of random noise to the initial state. Since the governing model equations of the atmospheric circulation define a deterministic non-periodic flow, error growth crucially depends on the local conditions of the model dynamics. Hence, random noise perturbations do not necessarily grow. This requires either an unstable layering of the atmosphere (convective zones), where the upscale error growth can be quite fast, or a self-organization of the noise into larger scale flow dependent patterns that drive the synoptic scale circulation. The latter process is much slower and may take some days (see e.g. \cite{bib:Selz2022}). 

There is an ongoing debate in the literature about the properties of upscale error growth from very local small amplitude random perturbations, which may originate from the ``wing beats of butterflies''. It is well known that NWP systems enable ``butterflies'' to disturb the larger synoptic scales. \cite{selz2023} found that this is not necessarily the case in MLWP. On the other hand \cite{bib:Durran2014} rank butterflies to be unimportant at all, because ``in any real-world event, the contributions of butterflies to uncertainties in initial conditions would be completely dwarfed by errors in the larger scales''.

Already in the 1990's the need for a more systematic exploration of the perturbation spaces in high dimensional dynamical systems motivated the implementation of a Singular Vector (SV) technique based on the Lanczos algorithm (L-SV) by \cite{bib:Mureau1993} and \cite{bib:Molteni1996}. For a comprehensive overview see \cite{bib:Diaconescu2012}. An example for the application in operational NWP can be found in \cite{bib:Leutbecher2008}. The assumption is that the sensitivities of a forecast model to small perturbations can be approximated by the Jacobians of the flow. In high dimensional systems, where the Jacobian is not available in matrix form, L-SV approximates the piece-wise tangent linear evolution of perturbations in a given optimization window by integrations with the tangent-linear and adjoint versions of the dynamic model. The result is a set of orthonormal flow-dependent perturbation vectors, which point in the directions of the fastest error growth. SVs are quite effective as noted e.g. by \cite{bib:Selz2022}. They state that ``the addition of singular vectors to the initial condition uncertainty (..) decreases the predictability time by about 1 day compared to the corresponding experiment without the singular vectors''. 

A disadvantage of L-SV is the need for setting up and maintaining linear and adjoint model versions what generates a huge additional effort. To address this shortcoming, we created an adjoint-free version of the SV method, which we call Arnoldi Singular Vectors (A-SV). The A-SV is based on forward integrations of the full nonlinear model only. For details, see \citep{winkler2020}.
 
In the present paper the Arnoldi-SV method is used with the Pangu Weather ML model of Huawei \citep{bi2023accurate} and we show that the algorithm constructs larger scale flow dependent perturbations, which alter the larger synoptic scales right from the beginning of the forecast.

This article is organized as follows: more details about the Pangu Weather model are given in section \ref{sec:Pangu-Weather}. We describe the mathematical and technical details of the Arnoldi-SV method in section \ref{sec:ArnoldiSV}. Section \ref{sec:noise_to_struc} compares the A-SV approach with Google-Deepmind's GenCast generic diffusion model \citep{price2024gencastdiffusionbasedensembleforecasting} and explains the conceptual differences in designing ensemble prediction systems (EPS). The final sections present the results and end up in concluding remarks. In the appendix (section \ref{sec:appendix}) we provide a larger amount of further information, in particular pseudocode algorithms of A-SV and additional plots.

\section{MLWP with Pangu Weather}\label{sec:Pangu-Weather}

One of the first MLWP systems which reached a forecast quality comparable to NWP was Pangu Weather from Huawei \citep{bi2023accurate}. This ML model implements a ``3D Earth-specific transformer'' NN architecture with encoder and decoder parts connected by a U-shaped processor. It is noteworthy that the degrees of freedom in the layers of the processor (latent space / bottleneck) are much smaller than the size of the processed atmospheric fields. From a mathematical point of view, this can be interpreted as an implicit decomposition plus cutoff that effectuate a data compression with denoising. 

The dampening of noise by Pangu Weather also relates to the model time step. While NWP models produce forecasts with very small integration time steps (seconds to minutes), MLWP models learn to connect atmospheric patterns separated by hours. For example, the team of Pangu Weather decided to set up four different Pangu Weather ML models predicting different lead times at 1h, 3h, 6h and 24h. The 24h model de facto concentrates on the long-living large scale synoptic flow and we use it throughout this paper. 

Pangu Weather is trained on the ERA5 re-analysis dataset \citep{hersbach2020era5}\footnote{ERA5 documentation: \url{https://confluence.ecmwf.int/display/CKB/ERA5}.} of ECMWF. The training dataset contains hourly atmospheric reanalysis states since 1940, where ECMWF had been running a 4DVar data assimilation cycle with a recent version of the NWP model from the Integrated Forecast System (IFS)\footnote{IFS documentation: \url{https://www.ecmwf.int/en/publications/ifs-documentation}}. The states have a horizontal resolution of 0.25 degree at 137 vertical pressure levels, while Pangu is trained on 13 selected levels. The quality of these NWP based 3D estimates of the atmospheric state has increased a lot with the upcoming use of satellite observations in data assimilation and ML models are usually trained with ERA data from 1979 onwards.

Surprisingly, the successful MLWP models use just a small subset of variables from an ERA5 system state. For example, Pangu Weather takes four surface variables (2m temperature, two 10m wind components, mean sea level pressure) and
five atmospheric variables (temperature, both wind components, specific humidity, geopotential) on 13 pressure
levels on input. Nevertheless, the processing of the still huge amount of data during training consumes large amounts of computation time. But once the model is trained, generating a forecast is very cheap and fast. The authors of Pangu Weather denote that one inference of the model is ``more than $10000\times$ faster than the operational IFS NWP model of ECMWF'' \citep{bi2023accurate}. 

\cite{selz2023} found a fundamental difference in error growth between ML models (Pangu Weather of Huawei \cite{bi2023accurate}) and NWP models, represented by the non-hydrostatic NWP model ICON of Deutscher Wetterdienst (\cite{bib:Zaengl2015}). While error growth in the ML model is independent from the size of the initial errors, the NWP model strongly reinforces small amplitude perturbations, that may originate from ``wing beats of butterflies'' \citep{lorenz1996essence}.

\section{Theory and Design of Arnoldi-SV}\label{sec:ArnoldiSV}

\subsection{Perturbation growth}

When integrating a perturbed initial state forward in time, the difference vector between the perturbed trajectory and the unperturbed reference changes length and direction, because some of the components grow while others shrink. Consequently, a random perturbation vector always turns into the subspace of growing errors. Given a suitable matrix with sufficient information on the local dynamics, the leading SV denotes the direction of largest error growth. Writing this down, allows us to see this claim more clearly (an analytical example for the Lorenz63 model is given in \cite{ExtendedRangeAtmosphericPredictionandtheLorenzModel}).

Let $x_0 \in \R^n$ be a reference state at initial time, 
$\delta_0 \in \R^n$ a small perturbation at the reference state and $\delta_{\tau} \in \R^n$ denotes the evolved perturbation at time point $\tau \in \R^+$. The matrix $A \in \R^{n \times n}$ has to be a suitable operator of the local perturbation dynamics around $x_0$ over the $\tau$-period of time, e.g. a tangent linear operator. Hence, the logarithm of the growth rate can be constituted as follows:
\begin{equation}\label{eq:SV1}
log \frac{||\delta_\tau||}{||\delta_0||} =
log \frac{||{\bf A} \, \delta_0||}{||\delta_0||} =
\frac{1}{2}
log \frac{({\bf A} \, \delta_0)^T {\bf A} \, \delta_0}{ \delta_0^T \delta_0} =
\frac{1}{2}
log \frac{ \delta_0^T \ {\bf A}^T {\bf A} \, \delta_0}{ \delta_0^T \delta_0}
\end{equation}
where $||.||$ denotes a suitable norm. The logarithm is chosen, because it relates to the popular concept of Ljapunov exponents in dynamical systems which quantify exponential error growth.

According to eq. \ref{eq:SV1} error growth can be categorized in terms of the eigenvectors of ${\bf A}^T {\bf A}$. The eigenvectors of ${\bf A}^T {\bf A}$ are equal to the (right) singular vectors of ${\bf A}$. The first mentioned is the equivalent eigenvalue problem, which is solved in the Lanczos SV algorithm (L-SV).
Our Arnoldi based approach, which we describe in the following sections in more detail, calculates the  singular value decomposition (SVD) directly of ${\bf A}$ without detour. 

\subsection{Krylov Methods}

In high dimensional systems it is usually not possible nor efficient to fully compute, represent and store an operator ${\bf A} \in \R^{n \times n}$ in matrix form. Krylov subspace methods (see e.g. \cite{bib:Golub1996}, \cite{bib:Meurant2020}) are a powerful group of numerical methods, which allows to handle such problems ``matrix-free''. This means that the algorithm does not need nor calculate the full matrix operator. Instead, only the ability to approximate matrix-vector products of ${\bf A}$ with an arbitrary vector $\bf v$ is required. A Krylov subspace of dimension $m$ is defined as follows:
\begin{equation}\label{eq:SV3}
{\kappa}_m({\bf A}, {\bf v}) = span\{{\bf v}, {\bf A} {\bf v},{\bf A}^2 {\bf v},...,{\bf A}^{m-1} {\bf v}\}.
\end{equation}
From this information the Arnoldi algorithm constructs an orthonormal Krylov basis ${\bf Q} \in \R^{n \times m}$ and a low-dimensional projection ${\bf H} \in \R^{m \times m}$ of the original operator ${\bf A}$. In particular, it holds that $m \ll n$ with $m,n \in \N$. The two operators ${\bf A}$ and ${\bf H}$ are linked by
\begin{equation}\label{eq:SV2}
 {\bf A}_{n,n}{\bf Q}_{n,m} = {\bf Q}_{n,m}{\bf H}_{m,m} +  {\bf \textbf{RES}},
\end{equation}
where {\bf \text{RES}} is the residual of the approximation and $[\cdot ]_{n,n}$ denotes in compact form that a matrix $[\cdot ]$ is part of the $\R^{n \times n}$ space. The assumption is that the projection of the SVs of ${\bf H}$ into the $n$-dimensional model space by multiplication with ${\bf Q}_{n,m}$ sufficiently covers relevant modes of ${\bf A}$, often already at early stages of the construction process.

\subsection{Dynamical Systems and Evolved Increments}

We denote the development of a (continuous) dynamical system by the following function
\begin{equation}
{\it M}_{\tau}: 
\R^n  \to \R^n, \ x \mapsto {\it M}_{\tau}(x).
\end{equation}
Here, ${\tau} \in \R_+$ is an arbitrary but fixed period of time, $x\in \R^n$ an state of the system and $n \in \N_+$ the dimension of the dynamical system. We define an evolved increment, as the difference of a perturbed and an unperturbed reference state after the specific time $\tau \in \R$. Let $\bf v \in \R^n$ be a normalized (perturbation) vector, ${\bf x}_0 \in \R^n$ a reference state and $h \in \R^+$ the perturbation amplitude. Now, the evolved increment $I_{\tau,{\bf x}_0,h}(v)$ writes as follows
\begin{equation}\label{eq:SV4}
I_{\tau,{\bf x}_0,h}({\bf v}) := {\it M}_{\tau}({\bf x}_0+ h {\bf v}) -{\it M}_{\tau}({\bf x}_0).
\end{equation}
We introduce the evolved increment matrix (EIM), which collects the evolved increments and is defined as follows:
\begin{equation}\label{eq:SV7}
{\bf A}_{\tau,{\bf x}_0,h} := \left( I_{\tau,{\bf x}_0,h}(e_1), \ldots , I_{\tau,{\bf x}_0,h}(e_n) \right) ,
\end{equation}
where $e_i \in \R^n$ denotes the $i$-th unitary vector. The optimization time $\tau$ is a parameter of the flux derived function $M_{\tau}$. Hence, the EIM defines a linear map, which approximates the observed error growth between initial time and the end of the optimization window.
We use the assumption that nonlinear evolved increments of arbitrary vectors can be approximated by a linear combination of the nonlinear evolved unitary vectors. This holds under the assumption of handleable regularity (smoothness) of the dynamical system in the neighborhood of the initial state and leads to the following connection:
\begin{equation}\label{eq:SV_approx}
I_{\tau,{\bf x}_0,h}({\bf v}) \approx \sum_{i=1}^{n} [ {\it M}_{\tau}( {\bf x}_0 + h \, {\bf e}_i ) - {\it M}_{\tau}({\bf x}_0) ] \, v_i = \left( I_{\tau,{\bf x}_0,h}(e_1), \ldots , I_{\tau,{\bf x}_0,h}(e_n) \right) \, {\bf v} = {\bf A}_{\tau,{\bf x}_0,h} {\bf v}.
\end{equation}
This legitimates the assumption, that a linear EIM, consisting of nonlinear evolved increments, is able to cover reliable information of error growth around one reference state. Moreover, it enables the approximation of the needed matrix-vector products for the Arnoldi runs
to obtain a low dimensional representation of the full problem, starting from an arbitrary initial vector. The leading SVs of an EIM will point in directions, where an integration with the full non-linear model generates the largest increments $I_{\tau,{\bf x}_0,h}$ after time $\tau$ (large singular values; unstable manifold) or tend towards (tail of the singular spectrum; stable manifold) the evolved reference state ${\it M}_{\tau}({\bf x}_0)$.  While
Krylov subspace methods are known for a good recovery of the leading parts (in particular in the case of eigenvectors)
they are not able to recover the tail of the spectrum at the same accuracy, since these modes are not dominating the local dynamics.

\begin{figure}[]
\centering
\includegraphics[width=9cm]{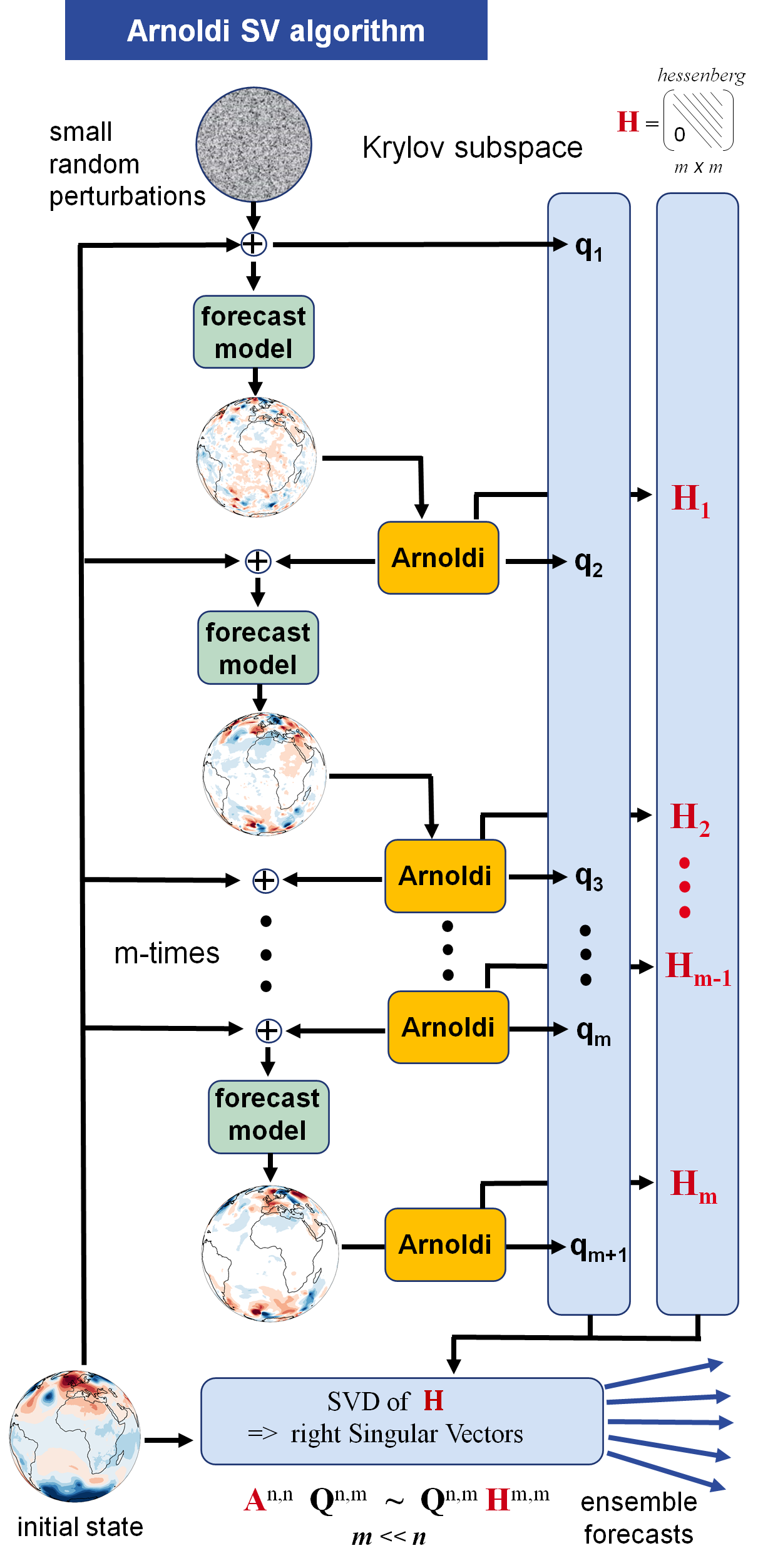}
\caption{A schematic representation of ensemble generation with the A-SV method. The algorithm (in the box) iteratively applies the forecast model to an initially random perturbation vector and creates a Krylov subspace and a basis thereof  ($Q=(q_1, ... , q_n)$) as well as the best representation of $A$ therein, which is $H$. Finally, a multiplication with $Q$ projects the SVs into the original space. The Krylov subspace is usually much smaller than the original space ($m<<n$).}
\label{fig:A-SV_perturbationConcept}
\end{figure}

\subsection{The Arnoldi-SV method}

The Arnoldi method was established by \cite{bib:Arnoldi1951}. A variety of versions exists in particular block versions, which can be run in parallel. Mathematically, the method is a generalization of the Lanczos algorithm \citep{bib:Lanczos1950} for non-symmetric matrices (see also \cite{bib:Saad1986}). 

Using the Arnoldi iteration in the context of dynamical systems is based on the EIM, which takes the evolved increments of nonlinear model runs. Fig. \ref{fig:A-SV_perturbationConcept} delineates the A-SV iterations for exploring the perturbation space at a given reference state. We initialize the algorithm with random noise perturbations. The perturbed atmospheric state is integrated forward in time with the full non-linear model. The evolved state is compared to the corresponding state of the reference trajectory at the end of the optimization window, what determines the evolved perturbation vector. From this information A-SV extends the operator $\bf H$ and constructs a new Krylov vector that is orthogonal to the existing Krylov subspace and is used to initialize the next A-SV iteration. After $m$ Arnoldi iterations the algorithm computes a SVD of the $m$-dimensional operator $\bf H$. The orthonormal Krylov basis ${\bf Q} = \{{\bf q}_i$, $i=1,2,...,m \}$ allows the projection of the SVs back to the $n$-dimensional model space according to eq. \ref{eq:SV2}. 

We provide two algorithms of the A-SV iteration in the appendix of this paper. A basic version (section \ref{ALG:asv-single}) and a block version (section \ref{ALG:asv-block}), which enables a block-wise computation of the Krylov vectors in parallel. 

A priori, it was not clear, if we could approximate SVs from an EIM. Hence, we used the term ``Block Arnoldi Perturbations'' (BAP) to distinguish these perturbations in our previous work. But we dropped this term, because the Arnoldi method indeed is able to find the growing modes of local perturbation spaces in dynamical systems.

For more details on the A-SV method and its mathematical plausibility we refer to \cite{winkler2020}, where we provide also error approximations.

\subsection{Properties of Arnoldi-SVs}

The A-SV method analyses the behavior of perturbations at the end of a given time window ($\tau$) just by comparing the perturbation vectors from both ends of this window and connects initial and evolved times by secants of the flow. On the one hand, this makes the method sensitive to the non-linear parts of error growth, but on the other hand ignores the evolution of the perturbations within the optimization window. The introduction of the EIM underlines that the strategy is not to construct the tangent linear flow thoroughly, but to find trajectories in model space that deviate from a given reference evolution.

In linear algebra a SVD, which consists of the left and right SVs and their singular values, is not related to any form of time evolution. The same is in general true for the Krylov iteration, since it does not distinguish any time points, there are just different vectors / states. But in the context of dynamical systems, the impact of the matrix operator $\bf A$ on a perturbation vector $\bf v$ is approximated by model forecasts (see eq. \ref{eq:SV_approx}), where $\bf v$ is added to the reference state at initial time (the onset of the optimization window) and the evolved perturbation $\bf A v$ corresponds to a state at the end of the window.

In principle, the A-SV method works for optimization windows of arbitrary length. But the longer $\tau$ gets, the more the nonlinear parts of the dynamics will dominate and the secants defined by $\bf H$ will become bad approximations of the perturbation dynamics. Hence, the length of the optimization window is an important parameter of the SV algorithms and one should cut longer optimization windows into appropriate pieces. This even holds for the A-SVs, although the problem is less strict compared to the tangent-linear Lanczos formulation, since the EIM is a matrix which consists of secants relying on non-linear forecasts. Due to these differences L-SV and A-SV will certainly set up different Krylov subspaces.

The amplitude is a free, but constant parameter of the evolved increment equation \ref{eq:SV4} and has to be chosen a priori. This choice is relevant and nontrivial. In particular, it depends on resolution, area, number of atmospheric parameter and level as well as the given norm. Since a common atmospheric data state consists of millions of grid points even very small local perturbations will result in a rather large numerical value for the amplitude of the perturbation vector, simply because of the large number of degrees of freedom. Hence, a suitable choice of an amplitude will rather be in the range of $100-10000$, than below $1$.
A value for a particular setup can be obtained by comparing two nearby model states with respect to the relevant area and variables, which represent initial condition error, (e.g. the evolved states of two short model forecasts with different lead times). We recommend to multiply this value by a factor in the range $[0.1,0.6]$. If a set of computed A-SV perturbations does not grow right from the beginning of the forecast, this is a relevant indicator that the amplitude was chosen too large.

\section{From Gaussian Noise to Structured Ensemble Perturbations}\label{sec:noise_to_struc}

Diffusion based ML gained wide attention in particular for image generation \citep{10081412}. Models like Dall-E, Midjourney or StableDiffuion came up only a few years ago and became very famous and powerful within a very short period of time. Diffusion models parametrize a Markov chain that gradually adds noise to the original data objects until the signal is destroyed (diffusion inference). ML models learn the reverse diffusion process (score based ``denoising'') to construct data objects from noise. New objects outside the training sample arise from the initialization of the ML model with a new set of random numbers (see e.g. \cite{bib:SongEtAl}, \cite{10.5555/3495724.3496298}, \cite{10.5555/3600270.3602196})

The reverse diffusion process can be guided to specific regions of the data manifold (conditioning) either by domain knowledge, where one selects and manipulates valid data objects or by setting more general rules and class descriptions that confine the target subspace. The latter is used for the most common approach, which are text-to-image generators, where diffusion models are combined with Large Language Models (LLM) in a way that text inputs are translated into rules for confining the data manifold.

Generic diffusion sampling was adopted by Google Deepmind's GenCast \citep{bib:Price2025} to set up ensemble prediction systems (EPS) in weather forecasting. The basic idea is that a successive sampling of forecast increments by a guided reverse diffusion process leads to a sequence of atmospheric states that will diverge from the reference trajectory. GenCast learns to construct 12h forecast increments conditioned on two consecutive ERA5 reanalysis states from noise. In each forecast time step GenCast randomly draws an increment by initializing the reverse diffusion sampling from a novel noise pattern. 

This has two distinct advantages: first, it enables the modeling of the forecast to some extend as a stochastic process, that randomly selects increments from the neighborhood of the ``true'' increment, where the neighborhood represents the total 12h forecast error of the ERA5 re-analysis dataset. This is a new way of modeling atmospheric dynamics. Second, the forecast rollout is less sensitive to the "regression to the mean" effect, because GenCast executes diffusion sampling in every forecast time step and the longest untouched forecast rollout time is 12h. Note, however, that GenCast may nevertheless smooth the forecast by dampening unskillful variability or processes with life times less than 12h.

As we have seen in the previous sections, the SV algorithms turn initial random noise into larger scale flow dependent patterns with distinct error growth properties. This is the result of the iterative application of the forecast model which defines a denoising filter based on the properties of the model dynamics. In MLWP SVs determine the sensitivities of a deterministic ML model to errors in the initial conditions. GenCast, however, not only learns to best predict ERA5 model states but also quantifies the forecast uncertainty involved. This includes model error, but also covers the intrinsic errors from the chaotic nature of the atmospheric circulation - which grow within the 12h forecast time window. 

While GenCast must sample $K$ stochastic trajectories to obtain a $K$ member ensemble, the A-SV approach needs to run only once at initial time and in principle produces a $K$ member ensemble from the $K$ leading SVs. Of course, the SV ensemble forecasts suffer from the ``regression to the mean`` effect in the same way as the unperturbed deterministic forecast. Setting up ML ensembles is beyond the scope of this paper and we will explore different options, including SVs, in future research.

\section{A-SV Experiments}\label{sec:sec-results}

\subsection{Configuration}

We compute SVs for the northern hemisphere at latitudes above $35$ degrees. We use the (massless) total energy described in section \ref{sec:total_eng} to measure distances between model states. We neglect the contribution from surface pressure, because it is by far the smallest contribution and it is known that kinetic energy (wind) covers the largest share of the total energy (e.g. \cite{bib:Diaconescu2012}, in particular figure 2 therein). We perturb the energy related variables $T,U,V$ on the $13$ pressure levels of the Pangu Weather input state and at the surface (2m temperature and 10m wind components). 

\begin{figure}[]
\centering
\includegraphics[width=13.5cm]{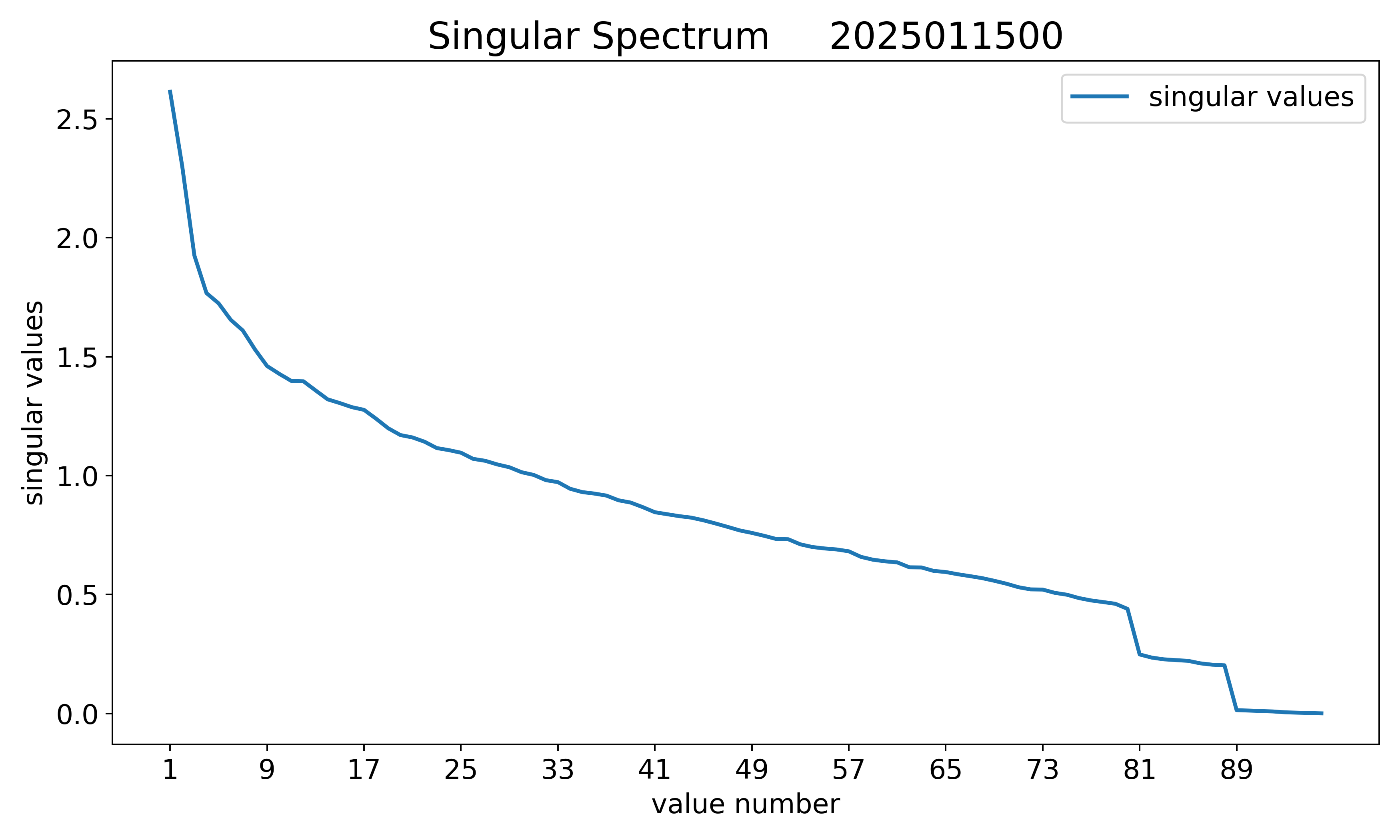}
\caption{Spectrum of obtained singular values. A-SV optimization period 24h, blocksize is 8 and number of loops 12. Reference state ICON $00$ UTC analysis state, $15$th January $2025$}\label{plot:singular_spec}
\end{figure}

The forecast inference is done with the 24h version of the Pangu Weather model and we set the length of the optimization window equal to this forecast time step of the model. We start the Arnoldi iteration always from Gaussian noise perturbations. We set the amplitude of the perturbations to $h=500$ throughout the iterations. We made that choice on the basis of observed variances of real states. We use the block A-SV Version (see section \ref{ALG:asv-block}) to enable parallel iteration and set the blocksize to $8$ and run $12$ loops. This means that we prepare $8$ different random noise patterns at initial time and obtain a final Krylov subspace of $96$ dimensions.

The presented results originate from daily A-SV experiments covering the period from Dec 2024 to Feb 2025. On each day the 00UTC global DWD-ICON analysis is used as the reference initial condition.

\subsection{Singular spectrum}

At the end of the A-SV iterations a Krylov subspace is generated and a SVD of the constructed low-dimensional operator $\bf H$ provides a singular spectrum. Figure \ref{plot:singular_spec} shows a typical spectrum. The number of singular values equals the size of the subspace ($96 =$ blocksize $8$ x $12$ loops), and they are plotted in descending order. We divide the matrix $\bf H$ by the amplitude $h$ to enable a more intuitive interpretation: singular values larger than one denote singular vectors of growing length, while values below one correspond to shrinking vectors. Note that this is only valid inside the Krylov subspace and can change when projecting the SVs into the full model space by multiplication with $\bf Q$. 

We would like to point out three things: First, the leading part of the spectrum decreases strongly with an outstanding leading singular value. This is common for real world problems. Second, somewhat less than one third of the values are above one denoting growing singular vectors (inside the Krylov subspace). Finally, the spectrum has two steps in its tail (between the 9th and 8th smallest values and the 17th and 16th smallest value), where the step size corresponds to the chosen blocksize of 8. While the step at the end of the spectrum is almost always in place, the second step in the tail occurs less often. These deformations of the spectrum indicate the information gains of the Krylov iterations. We recommend to use only the leading part of the spectrum for setting perturbations in ensemble forecasting.

\begin{figure}[]
\centering
\includegraphics[width=13cm]{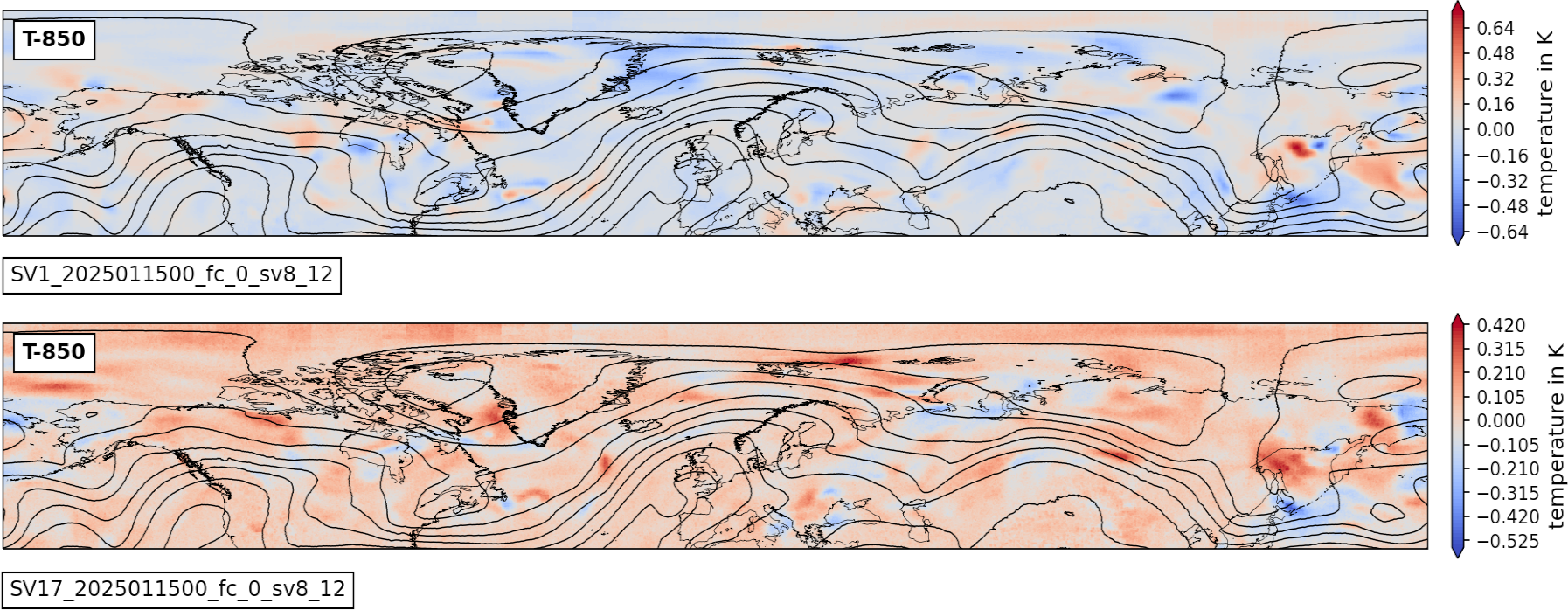}
\caption{A-SV perturbation 1.SV (top) and 17.SV (bottom), forecast leadtime 0h, temperature increment to reference 850 hPa and geopotential 500 hPa of reference state. Based on ICON $00$ UTC analysis state, $15$th Jan $2025$}\label{plot:SV1_SV17_temp}
\vspace{0.5cm}
\includegraphics[width=10cm]{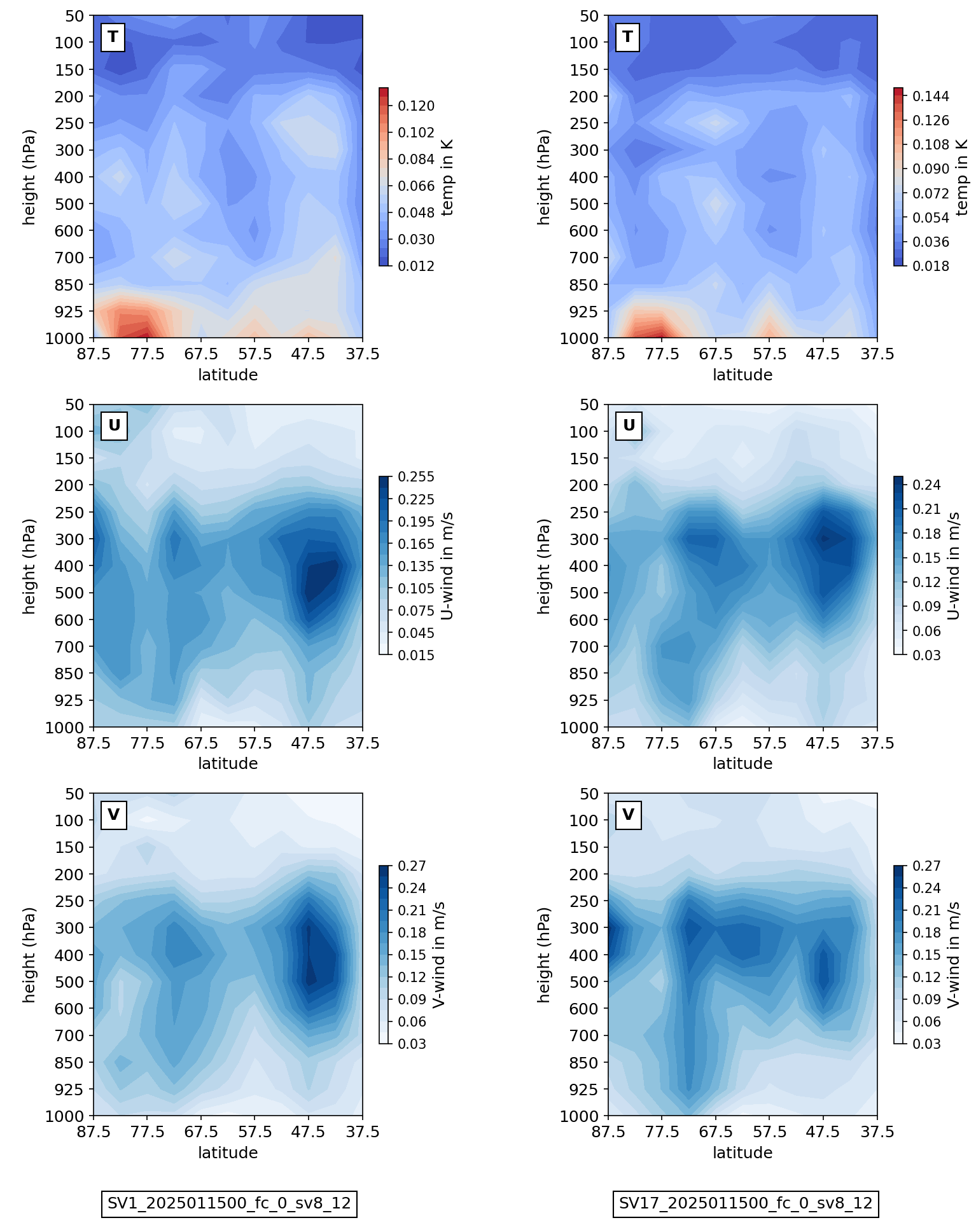}
\caption{A-SV perturbation 1.SV (left) and 17.SV (right), forecast leadtime 0h. Cross-section plots; average over latitudes of temperature (top row), U, V wind (middle row, bottom row resp.). Based on ICON $00$ UTC analysis state, $15$th Jan $2025$}\label{plot:SV1_SV17_css}
\end{figure}

\subsection{Structures of SV perturbations}

Figures \ref{plot:SV1_SV17_temp} and \ref{plot:SV1_SV17_css} compare the first and the 17th SV, where the latter is the first SV of the third block. The SVs are compared for the case study on 15th January 2025. Color shaded contours specify the perturbation patterns of the SVs and the solid isolines denote the $500$hPa geopotential of the reference state.  

The SV patterns modify the synoptic scales and seem to be located in sensitive areas of the atmospheric flow. Both shown SVs predominantly trigger different circulation structures.
The A-SV method is designed to set perturbations, where it gets the larger effects in terms of perturbation energy and this is indeed observable. These areas are near the surface for temperature and the jet streams for the wind components.

All SV are orthogonal. However, different SV can contain similar perturbation patterns, since mathematical orthogonality does not indicate that all structures appear completely different. Such similarities can be found in figure \ref{plot:SV1_SV17_temp}.

The so called cross-section plots provide wider insights in the vertical structure of the SV perturbations. For more information on this kind of plots see section \ref{sec:css}. Figure \ref{plot:SV1_SV17_css} shows cross-sections of the energy relevant variables T,U and V. We can see that the patterns of the temperature concentrate in areas near the surface. The wind components put the major part of the perturbation energy to the upper troposphere. This modifies the jet streams, where the SV algorithm can easily generate large amounts of kinetic perturbation energy already from small amplitude perturbations due to the large wind speeds and strong gradients at these atmospheric levels. The wind pattern at about 75 degree north, which we see for both SVs, stretches down to the lower troposphere and coincides with the increased temperatures in polar regions, what indicates a southward shift of the polar front. 
As we can see in figures \ref{plot:SV1_SV17_temp} and \ref{plot:SV1_SV17_css} both SVs trigger the same circulation structures, but they do it in a somewhat different way. 

\begin{figure}[bt]
\centering
\includegraphics[width=13.5cm]{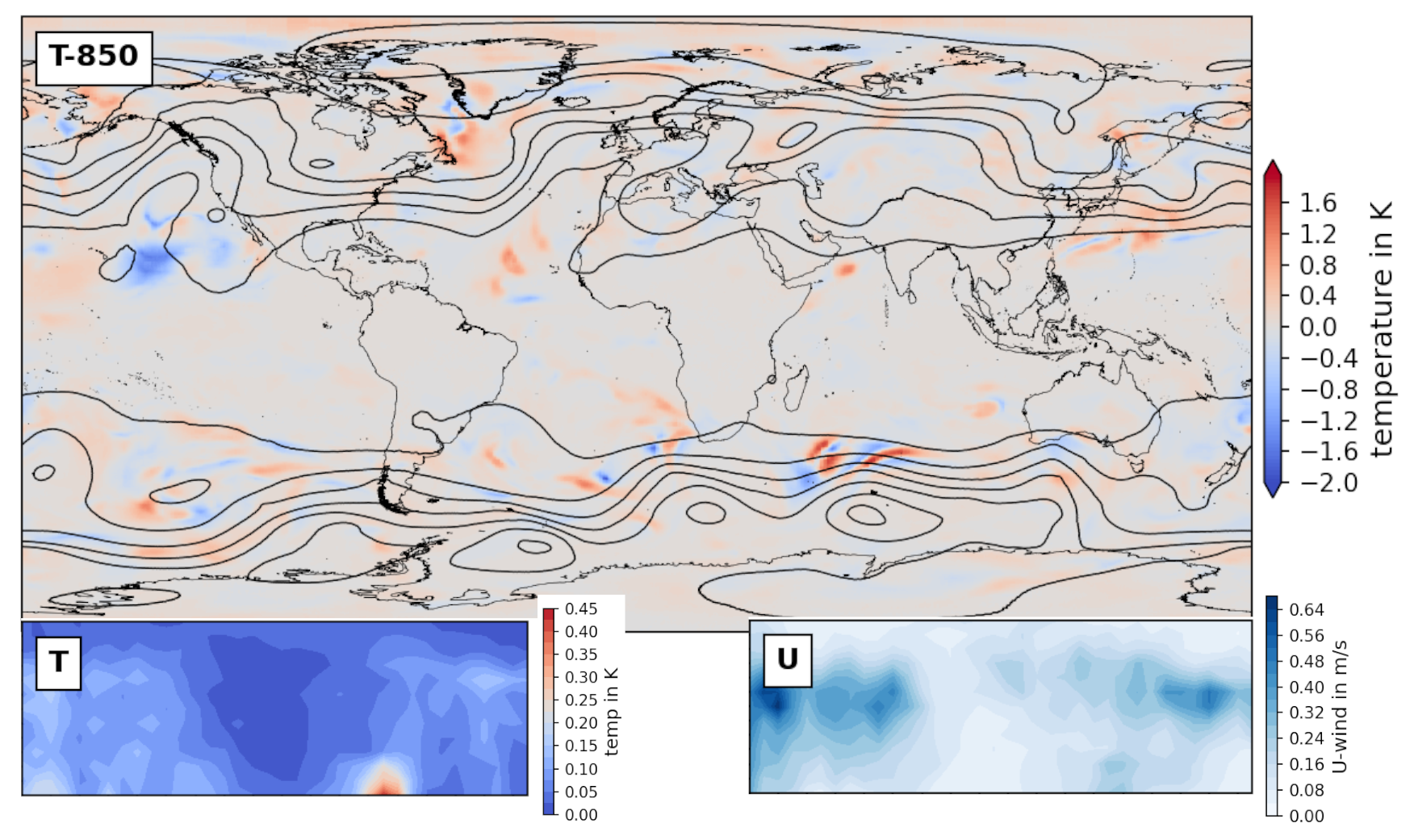}
\caption{A-SV perturbation 1.SV , forecast leadtime 0h, temperature increment to reference 850 hPa, geopotential 500 hPa of reference state (large plot). Cross-section plots over the whole earth (from north pole to south pole), of the temperature (bottom-left) and the U-wind component (bottom-right).  A-SV optimization period 24h, area whole earth, blocksize is 8 and number of loops 12 based on ICON $00$ UTC analysis state, $14$th January $2025$}\label{plot:SV1_global}
\end{figure}

Besides our main experiment for calculating local SVs on the northern hemisphere, we give a short insight to a global setup of A-SV, where figure \ref{plot:SV1_global} shows a case study from 14 January 2025. In this case the SV perturbation patterns concentrate on northern and southern hemisphere mid-latitudes and vanish from the tropics during the A-SV iterations for both wind and temperature. This occurs without additional restrictions and is the consequence of the fact that the A-SV together with the Pangu Weather ML model concentrate the perturbations to the dynamically active parts of the atmospheric circulation.

\subsection{Growth rates}

\begin{figure}[]
\centering
\includegraphics[width=13cm]{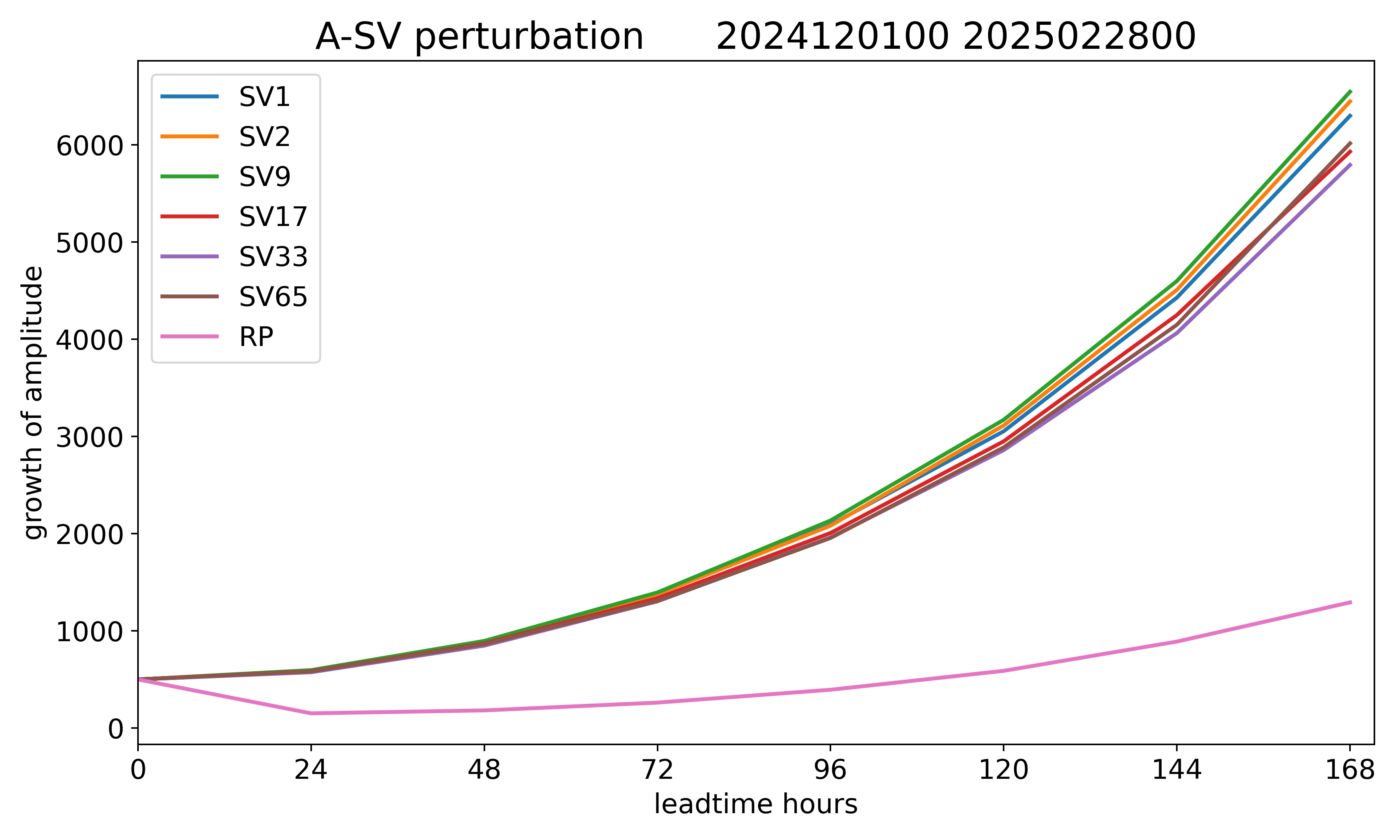}
\caption{Development of amplitudes of A-SV and random (gaussian noise) perturbations for forecasts of seven days / 168 hours. Optimization period 24h, blocksize is 8 and number of loops 12; daily runs based on ICON 00UTC analysis states, average over three months.}\label{fig:growth_of_amp}
\vspace{0.5cm}

\centering
\includegraphics[width=13cm]{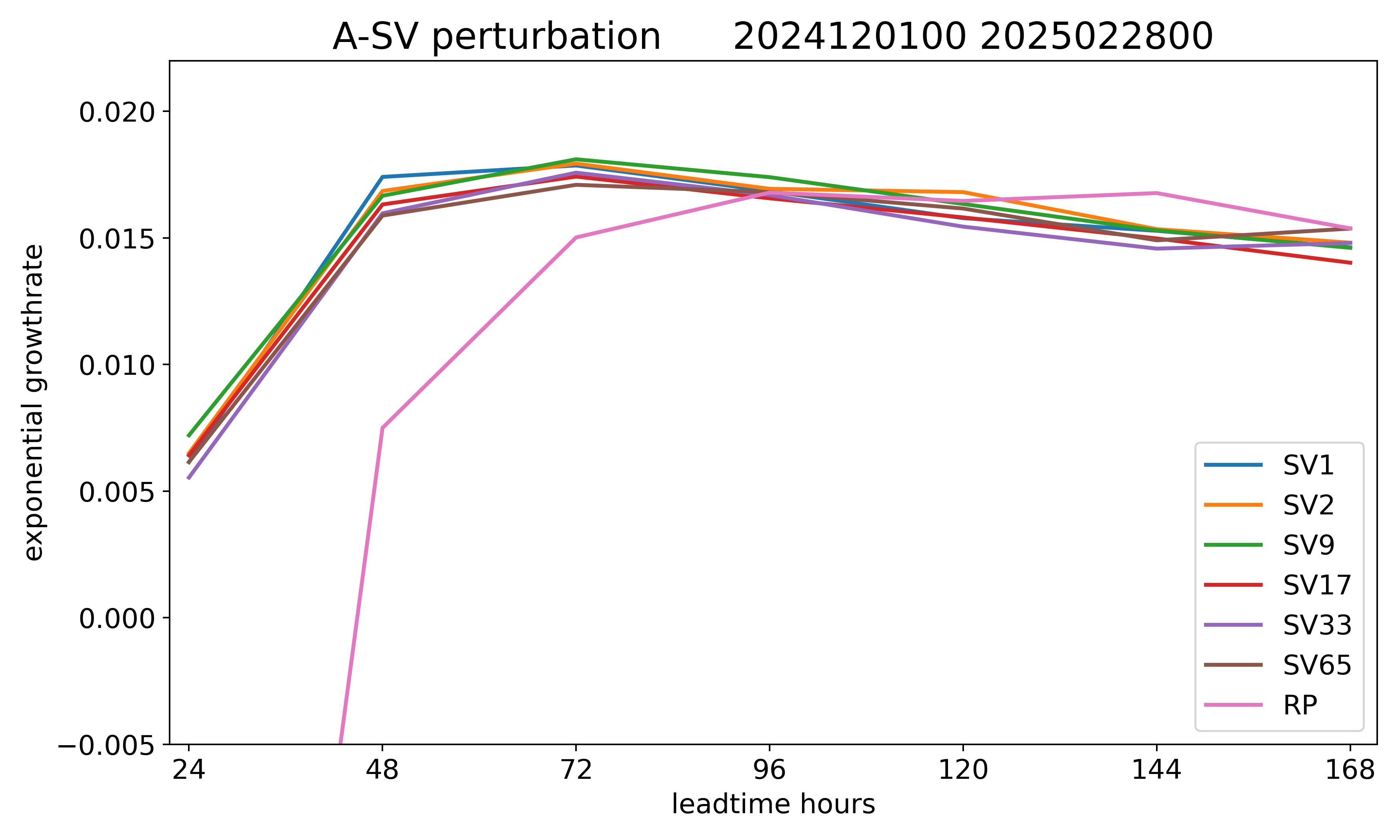}
\caption{Development of the mean exponential growth rate (see section [ref]) of A-SV and random (gaussian noise) perturbations for forecasts of seven days / 168 hours. Negative values denote shrinking, while positive values show growing behavior. Optimization period 24h, blocksize is 8 and number of loops 12; daily runs based on ICON 00UTC analysis states, average over three months.}\label{fig:growthrates}
\end{figure}

Figure \ref{fig:growth_of_amp} depict the absolute growth of perturbation amplitudes and figure \ref{fig:growthrates} provides mean exponential growth rates (MEGR) (see section \ref{sec:megr} for more details). The average amplitudes and growth rates are compiled from the three months experiment (Dec 24 to Feb 25). The pink solid line in the plots relates to random initial perturbations while all other lines denote error growth of selected SVs. These are the first two leading SVs and four selected SVs from different blocks. Both plots show the strictly different behavior of random noise and A-SV perturbations.

MEGR values measure the relative change in amplitude between consecutive forecast steps. They follow the concept of Ljapunov exponents (compare \cite{bib:bred-lyap}) and have been used by e.g. \cite{bib:magnusson_diss} to measure the perturbation growth in NWP. Negative values denote shrinking while positive values indicate perturbation growth. As can be seen in figure \ref{fig:growthrates} the absolute MEGR values are quite small, since usual growth rates in forecast models are limited in general. In particular growing perturbations concentrate in a few sensitive areas related to dynamically active circulation structures; in most places errors do not significantly grow.
Hence, this number range is more a property of forecasting models in particular the Pangu model, than one of the A-SV method.

Figure \ref{fig:growth_of_amp} provides an intuitive visual idea of the behavior of the perturbation amplitudes. It shows that the Pangu Weather ML model heavily dampens the amplitudes of random perturbations already in the first forecast time step, which corresponds to negative growth rates (see figure \ref{fig:growthrates}). It is plausible to expect such behavior from the neural network architecture of Pangu Weather, since the latent space of this model is much smaller than the dimension of the input state and causes an implicit filter, which eliminates unpredictable modes (see section \ref{sec:Pangu-Weather}). The remaining random perturbation energy organizes into larger scale patterns similar to those of the SVs (see sec \ref{sec:add_plots} in the appendix). Nevertheless, the amplitudes of the random perturbations remain small compared to those of the SVs, because these grow right from the beginning of the forecast. 

The finding that the random perturbations need about four days to recover the size of their initial amplitudes can be seen in connection with \citep{selz2023}. They showed that Pangu Weather has no ''butterfly effect'' and error growth is independent from the perturbation amplitude. Once the amplitude has decreased, there is no mechanism for compensating this deficit.      

A characteristic of chaotic dynamical systems is that perturbations can locally grow or shrink at different rates. However, in the limit of an infinitesimal time series with infinitesimal small time steps MEGR tends to a specific exponential growth rate, which is intrinsic to the system. This is the leading Ljapunov exponent. Indeed, figure \ref{fig:growthrates} shows a convergence of the (non-infinitesimal) MEGR, which might be used to characterize error growth for Pangu.

Figure \ref{fig:growth_of_amp} shows that the behavior of the SVs is pretty similar, but also uncovers a slight advantage of the three leading SVs (fig. \ref{fig:growth_of_amp}). The rather similar growth rates for different SV might result from the fact that the Krylov dimension in our A-SV runs is quite small compared to the size of the perturbed domain on the northern hemisphere ($~13$ million variables). Hence, the diversity inside the constructed Krylov subspace modes is limited. Nevertheless, Krylov methods are powerful algorithms, which are known for reasonable results even in small subspaces and - even more important - the process of construction is a priori not limited. Hence, it can cover arbitrary directions of the full model space in the constructed subspace. 
In summary, we can state that given suitable choices for the perturbed variables, domains and amplitudes, the A-SV approach is able to generate reasonable fast growing perturbations in the neighborhood of the reference initial state.

\begin{figure}[]
\centering
\includegraphics[height=5.5cm]{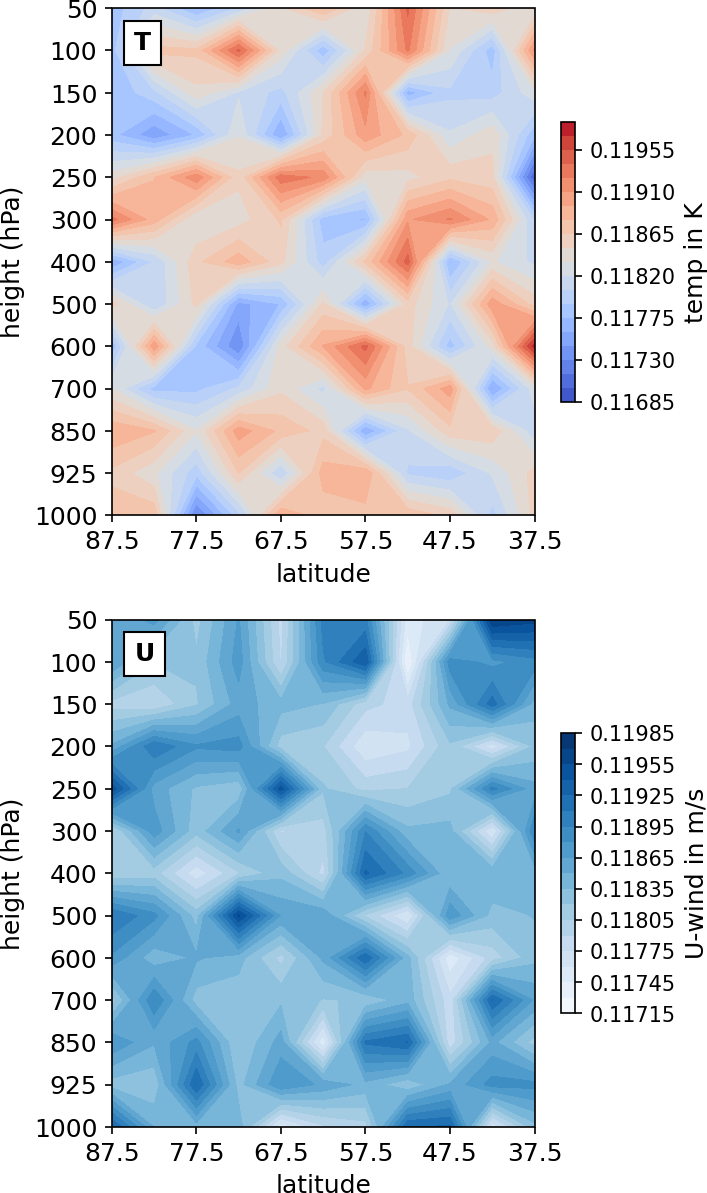}
\includegraphics[height=5.5cm]{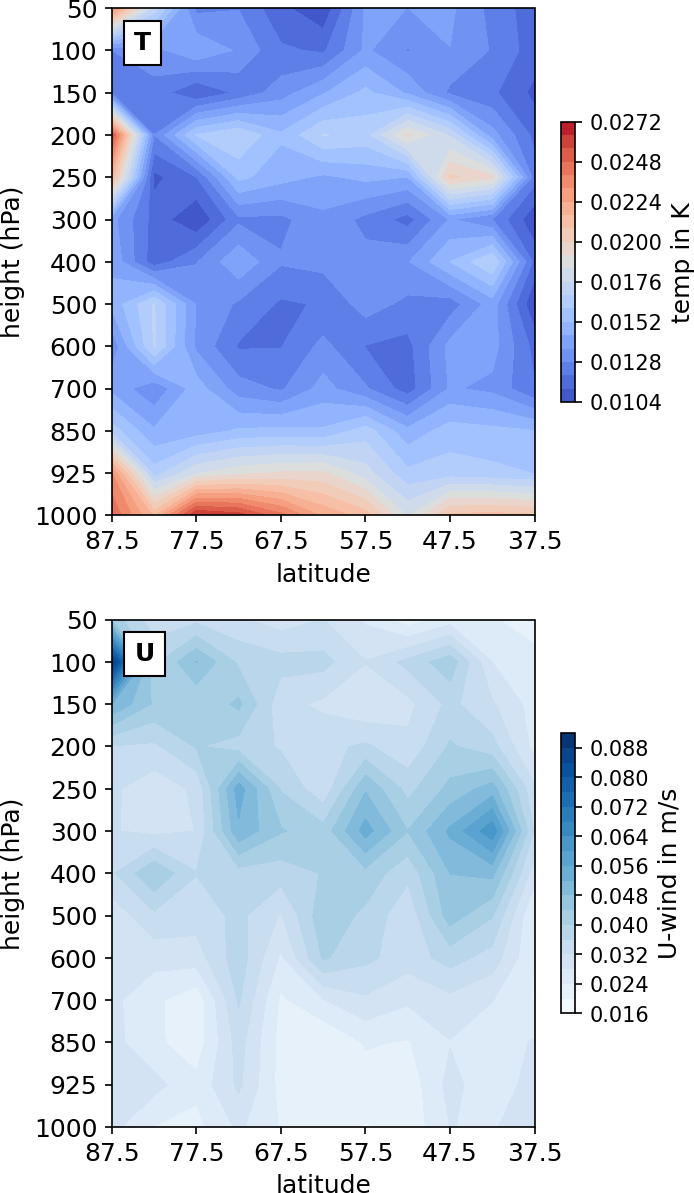}
\includegraphics[height=5.5cm]{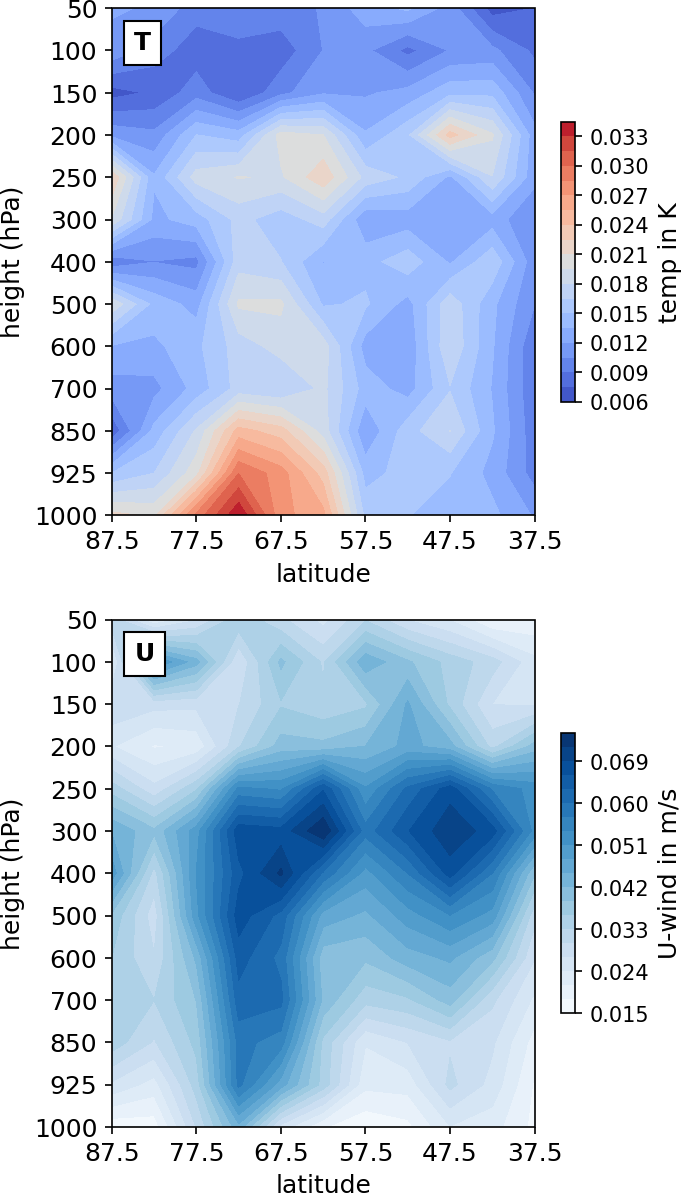}
\includegraphics[height=5.5cm]{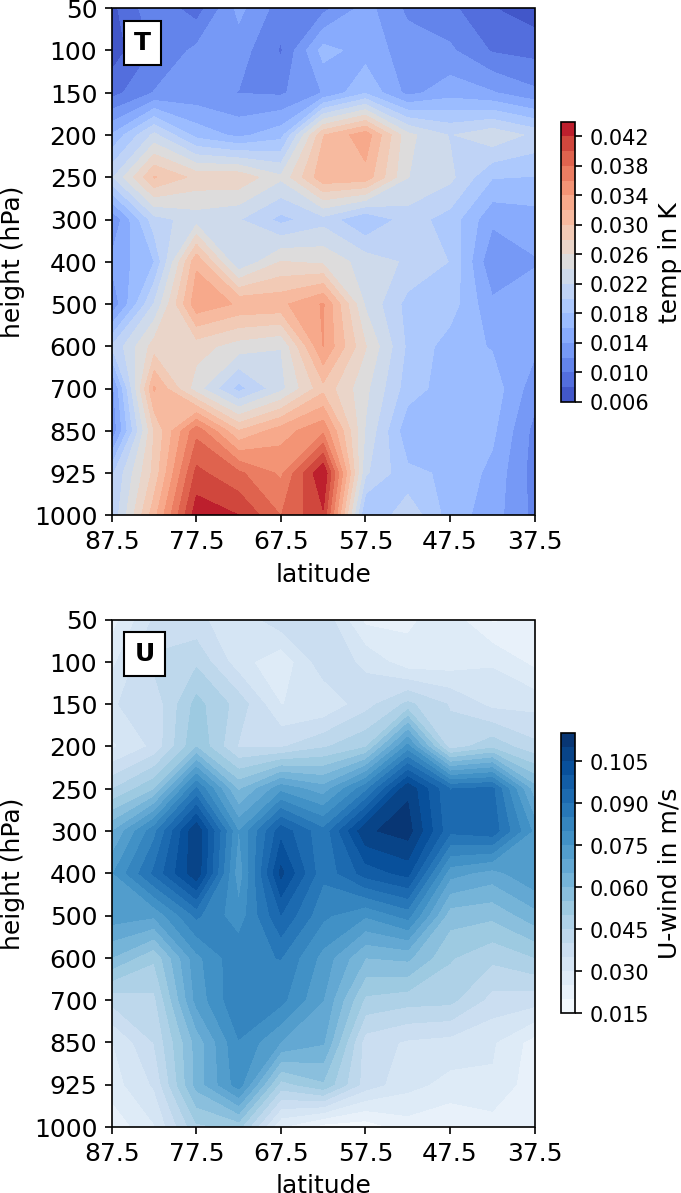}
\caption{Cross-sections of small random initial perturbations perturbations (RP) at latitudes >35 degree North for T (upper) and the U-wind component (lower). Generated perturbations (left) up to 72h lead time (right) in 24h steps. The perturbations are applied to the ICON $00$ UTC analysis on $15$th Jan. $2025$.}
\label{plot:crossSections_RP}
\vspace{0.5cm}
\includegraphics[height=5.5cm]{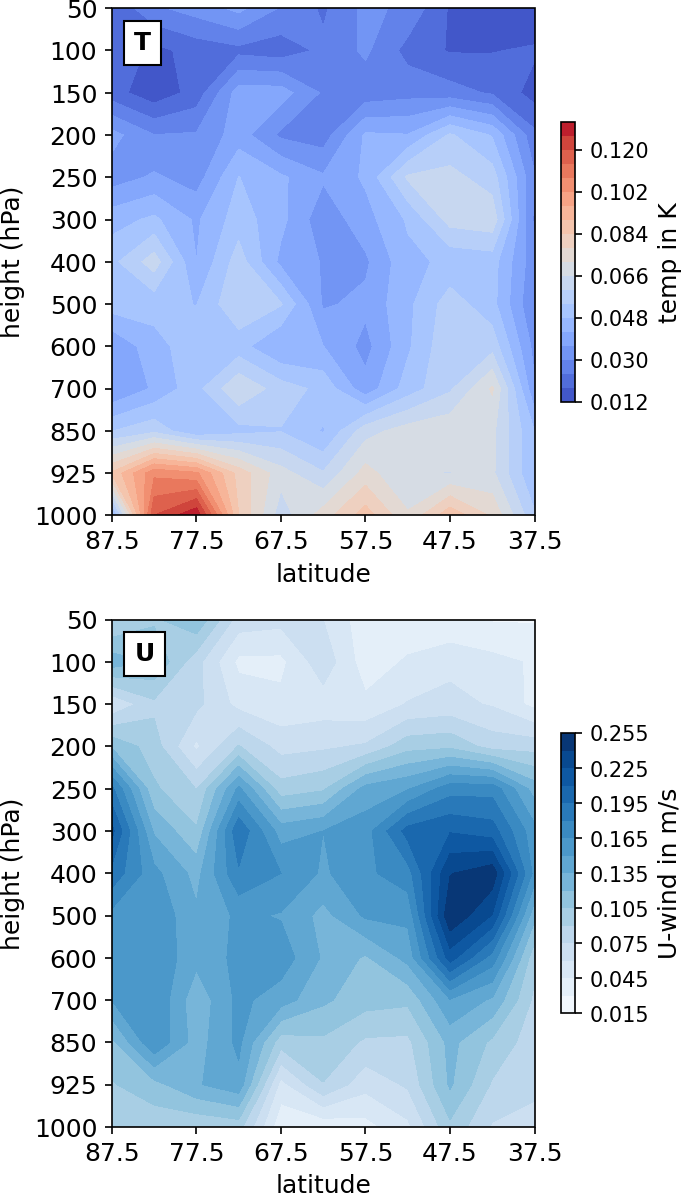}
\includegraphics[height=5.5cm]{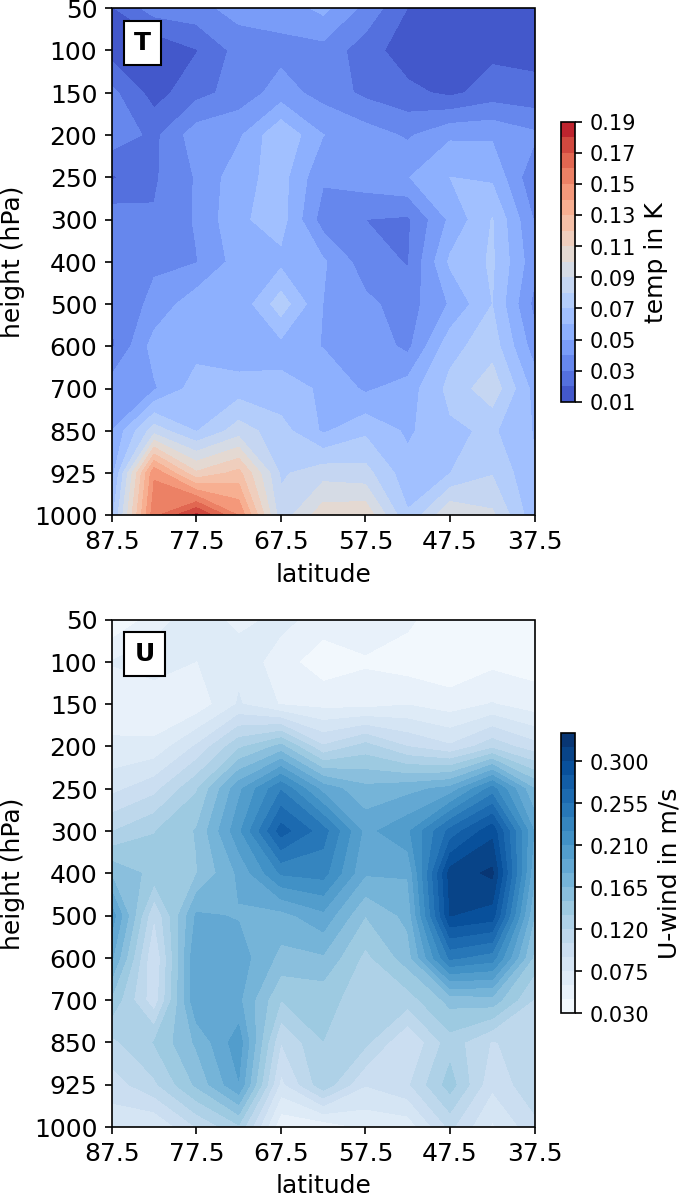}
\includegraphics[height=5.5cm]{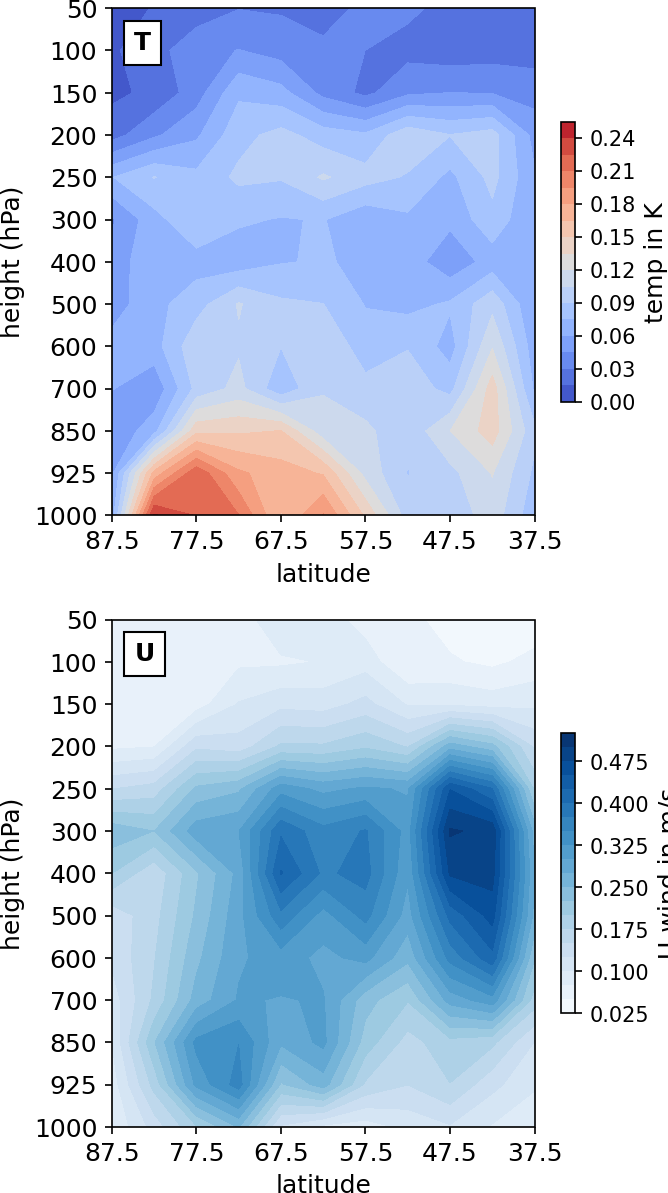}
\includegraphics[height=5.5cm]{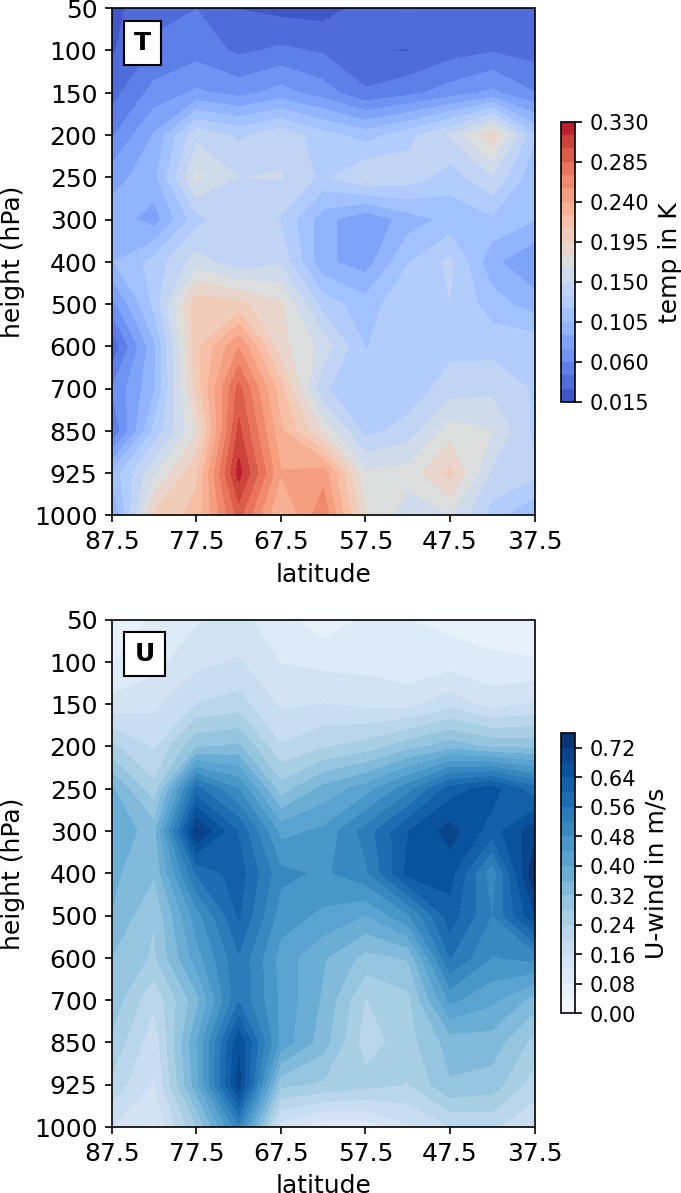}
\caption{same as figure \ref{plot:crossSections_RP}, but for SV1 }
\label{plot:crossSections_SV1}
\vspace{0.5cm}
\includegraphics[height=5.5cm]{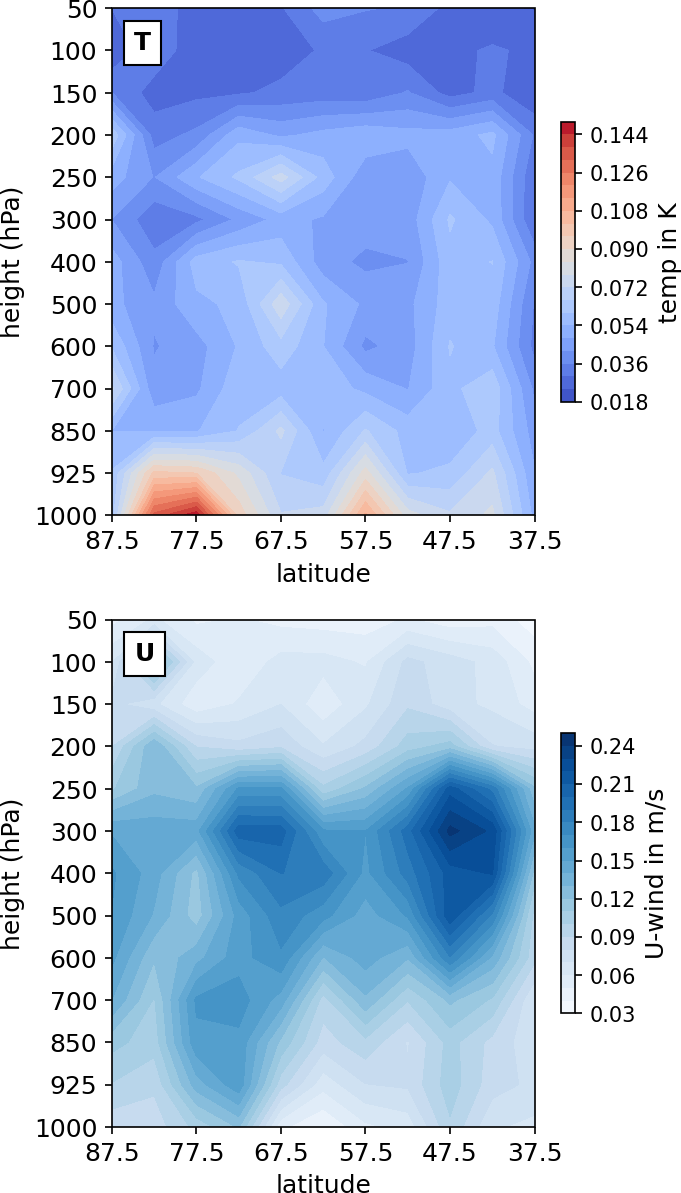}
\includegraphics[height=5.5cm]{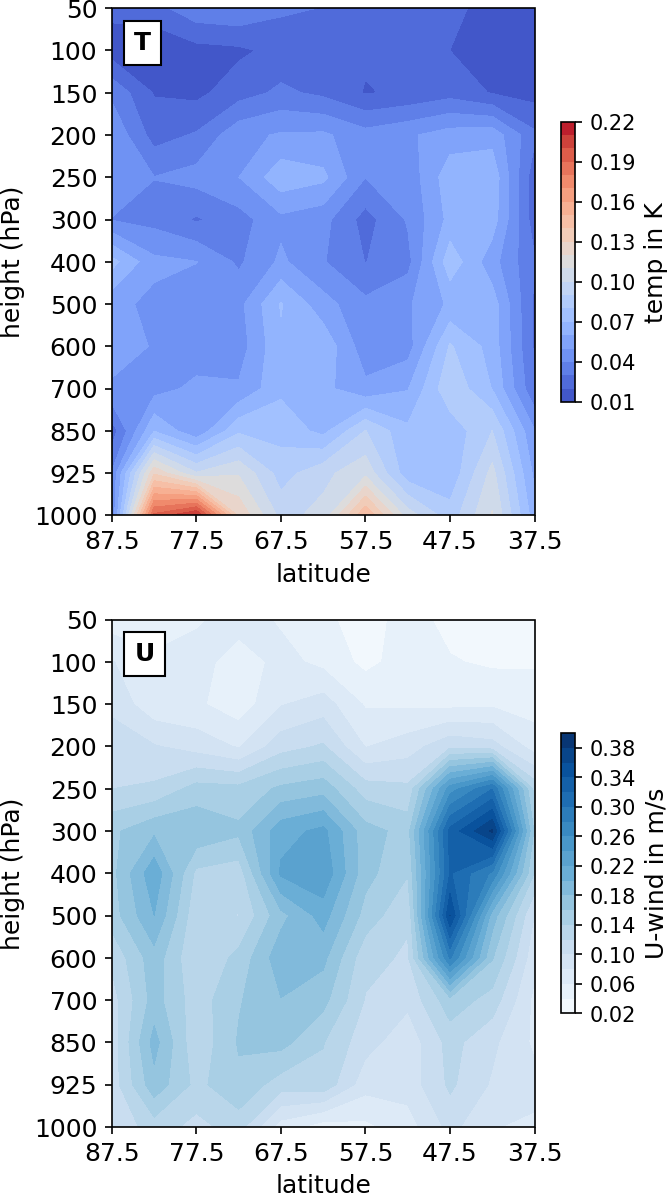}
\includegraphics[height=5.5cm]{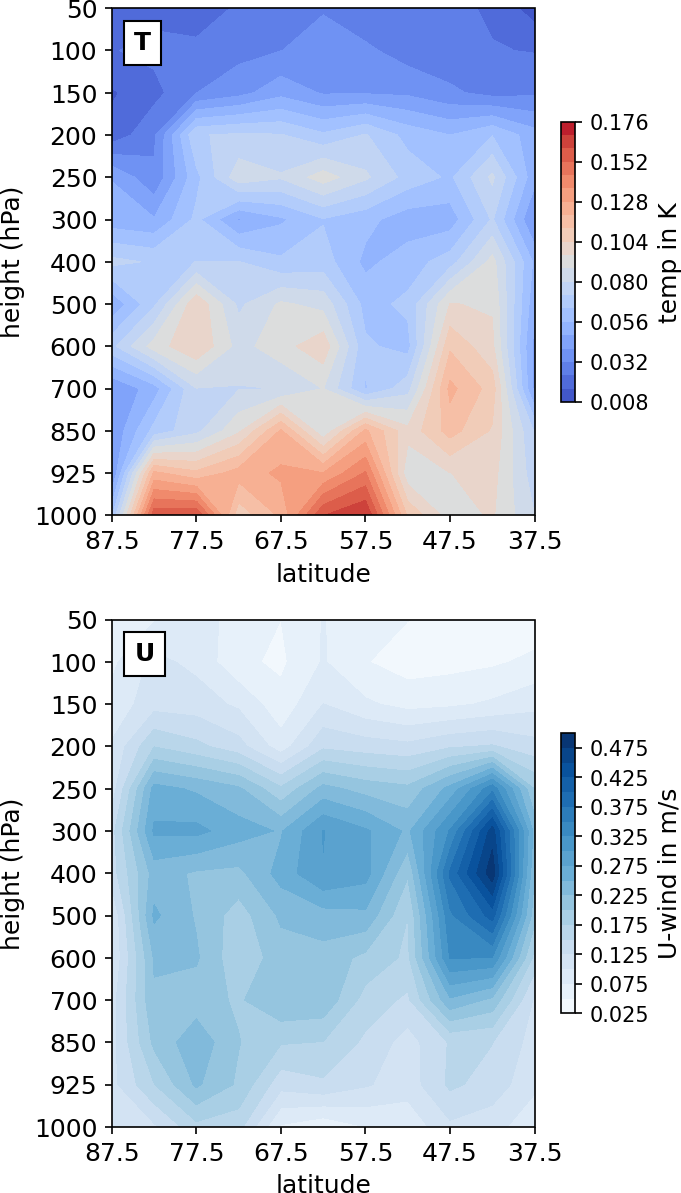}
\includegraphics[height=5.5cm]{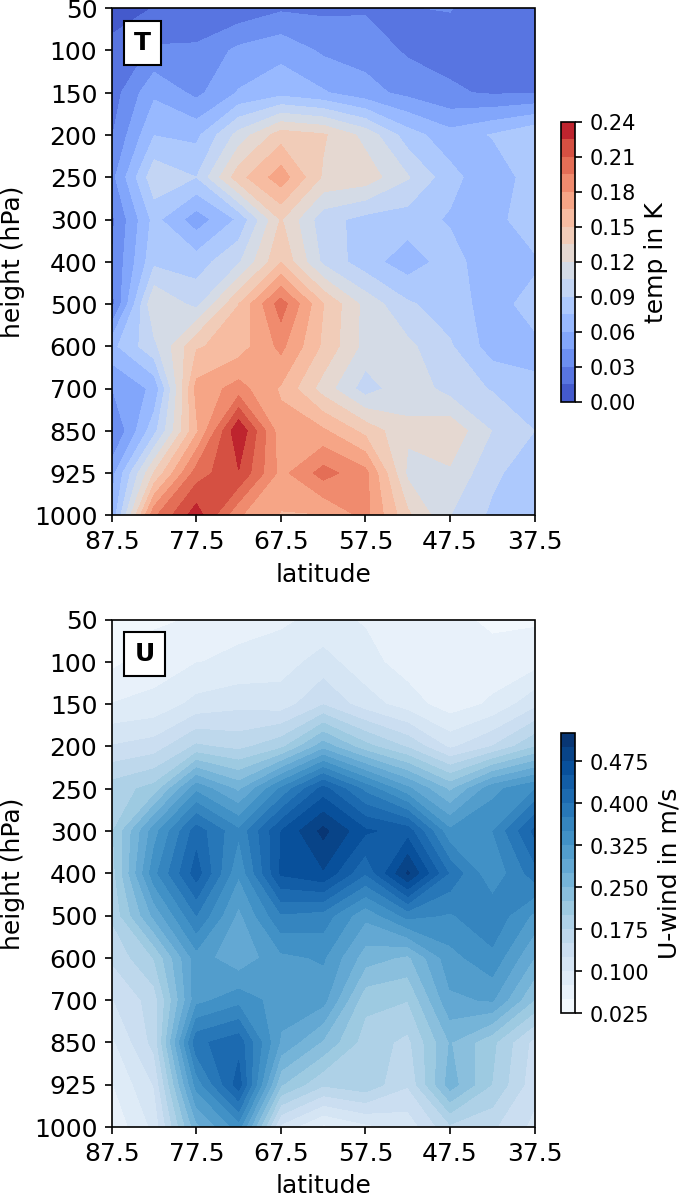}
\caption{same as figure \ref{plot:crossSections_RP}, but for SV17}
\label{plot:crossSections_SV17}
\end{figure}

\subsection{Evolution of perturbations in time}\label{sec:add_plots_dynamics}

The cross-section plots in figures \ref{plot:crossSections_RP} to \ref{plot:crossSections_SV17} demonstrate how small initial perturbations of the atmospheric circulation develop in the forecast. They show cross-section plots for latitudes above 35 degrees North for the temperature (T) and the U-wind component at forecast lead times up to 72h. The time difference between panels is 24h and the panels on the left show the perturbation patterns at initial time. Note that the range of the contour levels increases with lead time indicating growing perturbation amplitudes.

When simply adding noise to the initial atmospheric state of our shown case study ($15$th Jan. $2025$) we observed that the random patterns organize into flow-dependent perturbations already within the first time step of the 24h model at the expense of a large shrinkage of their amplitudes (fig. \ref{plot:crossSections_RP}). After 48h we see perturbation structures, which are rather similar to the SV perturbations in figures \ref{plot:crossSections_SV1} and \ref{plot:crossSections_SV17} at initial time (first panels on the left). But it takes about 96h until the random perturbation amplitudes return to their level at initial time (see previous subsection).

SV1 and SV17 start from perturbation patterns that look similar to those which develop from random perturbations at later stages of the forecast (see section \ref{sec:add_plots}). In contrast to the random perturbations in figure \ref{plot:crossSections_RP} their amplitudes start growing right from the beginning of the forecast (increased range of contour levels).

In all three cases, the temperature perturbations move to higher atmospheric levels in coincidence with the dislocation of the polar front.

\section{Summery, Conclusion and Outlook}\label{sec:Discussion}

We explored the dynamical properties of the pre-trained Pangu Weather ML model of Huawei with a specific kind of Singular Vectors (SV). An original version of the technique was introduced in NWP in the 1990's to find the perturbations of an initial state, which force the perturbed forecast to diverge from the reference. When applying SV algorithms to MLWP models we expect them to find those modifications of the input state that lead to significant errors at the output. More precisely, we expect that SV algorithms are suitable to find unstable / growing modes in the neighborhood of given reference states.

Our adjoint-free Arnoldi Singular Vector method (A-SV) observes error growth over a given optimization time window by applying the full non-linear forecast model to perturbed initial states. We start A-SV with Gaussian noise perturbations and found that it can generate flow dependent initial condition perturbations already from a small number of iterations. In contrast to random perturbations the A-SVs start growing right from the beginning of the forecast, which is the intended goal of the algorithm. 

As a next step, we will implement the A-SV method in an operational environment and explore its potential for generating reliable ensemble forecasts. One task here is to ensure that the sum of the initial perturbations add to zero. The easiest way to archive this, is a plus/minus paired approach, where one adds a perturbation vector twice with opposite sign. But it may be more flow dependent to implement a cluster approach, where the perturbations in the cluster are mixed and rescaled such that the cluster mean becomes zero.

Furthermore, we would like to point out that the introduced evolved increment matrix (EIM), can be combined with other algorithms from the family of Krylov Subspace methods, not just Arnoldi. This holds in particular for the Generalized minimal residual (GMRES) \citep{bib:saad_gmres} algorithm, which is used to solve large systems of linear equations, and we will explore this option in future research.

The most expensive part of the A-SV algorithm in terms of computational costs is the model forecast, while the matrix operations are relatively cheap. Therefore, the wide performance gain of MLWP compared to NWP is a major breakthrough. In combination with the block parallel version of the Arnoldi-SV method, the situation becomes even more comfortable and one might afford the re-construction of much larger Krylov subspaces than we generated in this paper. This can lead to a more complete approximation of unstable modes in MLWP systems and can help understand better the behavior of large MLWP models and enables interesting setups for Ensemble Prediction Systems (EPS).

\section*{Acknowledgement}

We would like to thank Berhard A. Schmitt and Stephan Dahlke from the working group Numerics and Optimization of
the Phillips Universität Marburg for their useful suggestions and help in preparing this article. We also thank the
BMV - Forschungsnetzwerk (German Federal Ministry of Transportation - Network of Research) for the financial support.

\bibliography{literatur_MLSV}

\newpage

\section{Appendix}\label{sec:appendix}

\subsection{Total Energy Measurement}\label{sec:total_eng}

Singular Vectors depend on the norm that is used to measure distances in the state space of the forecast model. In the field of weather prediction these are atmospheric states defined by different physical quantities like temperature or wind speeds. Hence, the euclidian norm might not be optimal, due to the different units and ranges of the physical quantities. In the literature different norms have been discussed, where the total energy seems the most natural choice. The measurement of (dry) total energy can be found in \cite{ehrendorfer1995} and is given by
\begin{equation}\label{eq:TEM}
|| x ||_E := \frac{1}{2} \int_{p_0}^{p_1} \int_S \left( U^2 + V^2 + \frac{c_p}{T_r} T^2 \right) \text{d}s \ \ \text{d}p \ +
	\frac{1}{2} \frac{R_d T_r}{p_r}  \int_S {ps}^2  \text{d}s
\end{equation}
where $p_0, p_1 \in N$ are the minimum and maximum pressure levels, respectively, $S$ the (horizontal) area of interest (e.g. the sphere) and $x=(U,V,T,ps)$ the atmospheric state consisting of the (horizontal) wind components, the temperature and the surface pressure, respectively. Furthermore, $T_r=270K$ denotes the reference temperature, $c_p= 1005.7 J kg^{-1} K^{-1}$ the specific heat capacity at constant pressure, $R_d=287.04 J kg^{-1} K^{-1}$ the gas constant of dry air
and $p_r= 1000 hPa$ the reference pressure. A discrete version can be defined as follows:
\begin{equation}\label{eq:DTEM}
|| x ||_{E} := \frac{1}{2} \sum_{j=1}^{p_n} \sum_{i=1}^{s_n} \left( U_{i,j}^2 + V_{i,j}^2 + \frac{c_p}{T_r} T_{i,j}^2 \right) \Delta s \ \ \Delta p \ +
	\frac{1}{2} \frac{R_d T_r}{p_r}  \sum_{i=1}^{s_n}  ({ps}_i)^2  \Delta s
\end{equation}
with analog meaning of the variables. 

The energy in eqs. \ref{eq:TEM} and \ref{eq:DTEM} are often denoted as ``(dry) total energy norm''. But from a mathematical point of view, the two equations do not define a valid norm, because therefor the axiom of absolute homogeneity ($||\lambda x|| = \lambda ||x||$, for some scalar $\lambda$) has to hold, which is obviously not true in both cases. We prefer to denote it mathematically more correctly as ``total energy measurement'' (TEM). 

The Arnoldi cycle requires computations of several scalar products. In order to keep the method consistent, a scalar product must imply the used norm. More mathematically spoken: We need a Hilbert space, where the scalar-product introduced norm corresponds with the TEM. Therefore, each atmospheric state $x=(U,V,T,ps) \in \R^n$, can be transformed into an energy consistent state vector $\hat{x}=(\hat{U},\hat{V},\hat{T},\hat{ps}) \in \R^n$, which is defined by
\begin{equation}\label{eq:ESV1}
\hat{U}_{i,j}~:=~\sqrt{ \frac{1}{2} \Delta s \Delta p }\ U_{i,j}, \,
\hat{V}_{i,j}~:=~\sqrt{ \frac{1}{2} \Delta s \Delta p }\ V_{i,j},  \,
\hat{T}_{i,j}~:=~\sqrt{ \frac{1}{2} \Delta s   \Delta p  \frac{c_p}{T_r}}\ T_{i,j}, \,
\hat{ps}_{i} := \sqrt{ \frac{1}{2} \Delta s \ \frac{R_d T_r} {p_r} }\ {ps}_{i}
\end{equation}
for each of the energy relevant atmospheric variables. $\Delta s$ and $\Delta p$ denote the discrete mesh size of the horizontal mesh and the vertical levels, respectively. If we transform these quantities, the relative ratio of two distances remain unaltered. We also choose $\Delta s = \Delta p = 1$. Based on these definitions the equality
$<\hat{x},\hat{x}> = ||\hat{x}||_2^2 =  ||x||_E$ 
holds. The Euclidean norm based on the standard scalar-product in the ``energy space'' now measures the square root of the total energy of a state. The root function is strictly monotonic and therefor provides a consistently measurement of the total energy. While this is a suitable solution, we like to point out that running Arnoldi with TEM, includes several transformations between the original space and the ``energy space''.

Since surface pressure is in real world approaches by far the smallest part of the total energy (e.g. figure 2 in \cite{bib:Diaconescu2012}) and it is rather expensive to handle, we do not perturb surface pressure and do not use it for calculating the total energy. Hence, we used a slightly reduced variant of TEM.

\subsection{Arnoldi-SV Algorithm}\label{ALG:asv-single}

We provide an algorithm of the A-SV approach.

\textbf{Arnoldi-SV Algorithm}\\[-8mm]
\begin{algorithmic}[1]\label{alg:ASV}\Statex
\Procedure{arnoldi-sv}{$\tau,h,x_0,q_1,m,\hat{k}$}
\Statex $\tau \in [0,\infty)$ \Comment{optimization time}
\Statex $h \in [0,\infty)$ \Comment{amplitude of perturbation}
\Statex $x_0 \in \R^n$ \Comment{unperturbed state of the system}
\Statex $q_1 \in \R^{n}$\Comment{initial vector}
\Statex $m \in \N$\Comment{number of loops within Arnoldi algorithm}
\Statex $\hat{k} \leq m  \in \N$\Comment{number of SV to be returned}
\Statex

\For{$j=1 :  m$}

\State $w \gets I_{\tau,x_0,h} \left( q_j \right) $\Comment{$w \in \R^{n}$ }

\For{$i=1:j$} \Comment{Gram-Schmidt orthogonalization}
\State $H_{[i,j]} \gets q_i^T\ w$

\State $w \gets w-q_iH_{[i,j]}$
\EndFor

\State $q_{j+1} \gets w$\Comment{new Krylov basis vector}
\State $H_{[j+1,j]} \gets ||w||$

\EndFor

\State $[U ; \Sigma ; V ] \gets \text{svd} \left( H \right)$ \Comment{singular value decomposition}
\For{$i=1:\hat{k}$}
\State $P_{i} \gets Q V_{i}$  \Comment{$Q = (q_1, \ldots , q_m) \in \R^{n\times \ell m}$}

\EndFor

\State \textbf{return} $P$ \Comment{$P \in \R^{n\times \hat{k} m}$}

\EndProcedure

\end{algorithmic}
\textbf{end}

The matrix $P$ contains the leading $\hat{k}$ SV approximations.
One may divide the obtained Matrix $H$ by the amplitude $h$ afterwards or as a variant may expand the algorithm with such a division. Then the matrix and the resulting singular values will be relative to the initial amplitude, which allows a better interpretation. See section \ref{sec:sec-results} for more information therefore. Please note, that if the A-SV algorithm should be used with TEM, several transformations between physical and energy space are required, which are not mentioned in the provided algorithms here.

This algorithm is a non-block (blocksize equals one) version to compute A-SV,
but Arnoldi iteration exists in different variants. Besides the standard version, where Krylov basis and matrix $H \in \R^{m \times m}$ are created step by step, the block versions of Arnoldi are of special interest. Instead of only one forecast per loop, the block versions allow to run the iteration with $l \in \N$ (blocksize) forecasts per loop \textit{in parallel} and the subspace grows by $l$ dimensions in each iteration. This makes the algorithm much more attractive, because a parallelization allows a large speedup. We provide a block version of the Arnoldi-SV algorithm in the following section.

\subsection{Block Arnoldi-SV Algorithm}\label{ALG:asv-block}

To compute Arnoldi-SV in parallel, a block version needs to be used. There exist different block versions of Arnoldi and in most cases these use a QR decomposition for the orthonormalization of one block. But we decided to use Ruhes Block Variant of Arnoldi (Alg. 6.24, \citep{bib:saad2003}, following \citep{bib:ruhe79}) instead. This version contains two properly nested loops in order to obtain an orthonormalization of the blocks by the usual Gram-Schmitt procedure. Since Gram-Schmitt-orthonormalization is very easy to implement and furthermore already part of the standard Arnoldi iteration, this allows an easy implementation of a block version of Arnoldi.\\

\textbf{Ruhe Block Arnoldi SV Algorithm}\\[-8mm]
\begin{algorithmic}[1]\Statex
\Procedure{block-arnoldi-sv}{$\tau,h,x_0,Q,m,\hat{k}$}
\Statex $\tau \in [0,\infty)$ \Comment{optimization time}
\Statex $h \in [0,\infty)$ \Comment{amplitude of perturbation}
\Statex $x_0 \in \R^n$ \Comment{unperturbed state of the system}
\Statex $Q \in \R^{n\times\ell}$\Comment{$\ell$ initial vectors merged in a matrix}
\Statex $m \in \N$\Comment{number of loops within Arnoldi algorithm}
\Statex $\hat{k} \leq m\ell\in \N$\Comment{number of SV to be returned}
\Statex

\State $Q \gets \text{orth}(Q)$ \Comment{orthonormalization}
\State $j= \ell$
\While{$j<=  m$}
\State $\kappa \gets [j-l+1 : j] $\Comment{$\kappa \in \N^{1 \times \ell }$ }

\For{$r=1:\ell$} \Comment{parallelizeable}
\State $W_{r} \gets I_{\tau,x_0,h} \left( Q_{\kappa_r} \right) $\Comment{$W \in \R^{n \times\ell}$ }
\EndFor

\For{$r=1:\ell$}
\For{$i=1:j$} \Comment{Gram-Schmidt orthogonalization}
\State $h_{i,\kappa_r} \gets \ <Q_{(i)}\ , W_r >$

\State $W_r \gets W_r - Q_{(i)} h_{i,\kappa_r}$
\EndFor

\State $ h_{j+1,\kappa_r} \gets ||W_r||_2$
\State $ Q_{(j+1)} \gets W_r * \frac{1}{h_{j+1,\kappa_r}}$

\State $j = j+1$
\EndFor
\EndWhile

\State $[U ; \Sigma ; V ] \gets \text{svd} \left( H \right)$ \Comment{singular value decomposition}
\For{$i=1:\hat{k}$}
\State $P_{i} \gets Q V_{i}$  \Comment{$Q  \in \R^{n\times\ell m}$}

\EndFor

\State \textbf{return} $P$ \Comment{$P  \in \R^{n\times \hat{k}}$}

\EndProcedure

\end{algorithmic}
\textbf{end}

where $P$ contains the leading $\hat{k}$ SV Perturbations. \



\subsection{Perturbation growth}\label{sec:megr}

We define an exponential growth rate (EGR) as follows
\begin{equation}
\text{EGR}_{x_0,\Delta t} :\R^+ \to \R , \ t \mapsto
 \frac{1}{\Delta t} \log \left( \frac{||M_t(x_0+hv) - M_t(x_0)||}{||M_{t-\Delta t}(x_0+hv) - M_{t-\Delta t}(x_0)||} \right).
\end{equation}
Here, $x_0$ is an unperturbed state of the system and $hv \in \R^n$ is a perturbation (scaled with amplitude $h \in \R+$). The time variable is denoted by $t$, $\Delta t$ is the used discrete timestep, the underlying dynamical system is defined by $M$ and $M_t$ defines an evolution over $t$ timeunits. We use the mean of several EGR in order to obtain a mean exponential growth rate (MEGR) (see \cite{bib:magnusson_diss}).

\subsection{Cross-section Plots}\label{sec:css}

\begin{wrapfigure}{r}{6.9cm}
\centering    
\includegraphics[width=3.5cm]{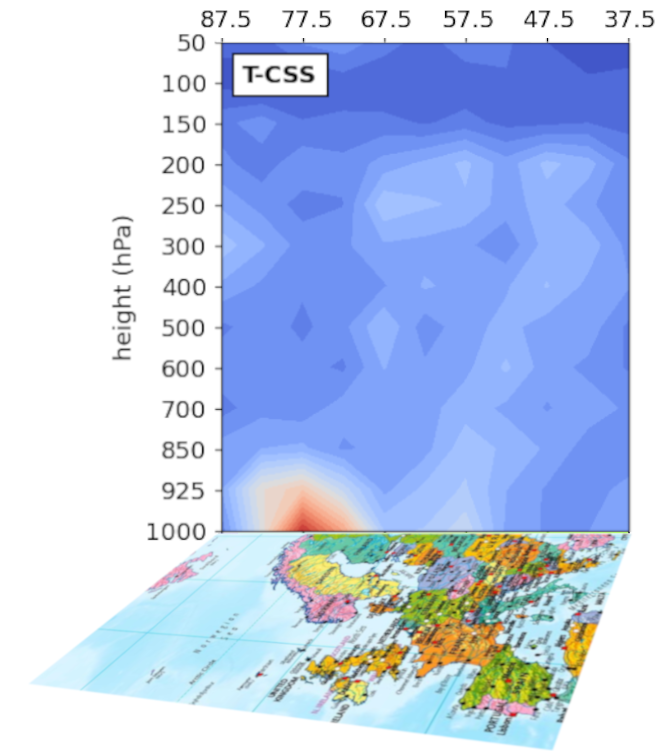}\label{plot:css_example}
\end{wrapfigure}

Cross-section plots enable to show the vertical structure of atmospheric states instead of only showing one level. The shown values are the mean of the absolute values along the latitudes on the different vertical pressure levels. We take the absolute values, because otherwise positive and negative amplitudes along specific latitudes might extinguish. Hence, cross-section plots are sort of a ``compression of the atmosphere'' along the latitudes. The figure on the right gives a visual impression of the idea, in that case over Europe.


\subsection{Additional plots}\label{sec:add_plots}

In order to provide more information of generated perturbations and their development over time, we provide a larger amount of additional plots.

This section contains figures of states with several variables, layers as well as cross-sections (see section \ref{sec:css}). 
The large plots show the unperturbed mean sea level pressure (PMS), temperature on 850 hPa and kinetic energy at 300 hPa. 
Please note that since we use a massless kinetic energy, the measurement of the occurring values is not physical and should there not interpreted in a strict physical way, but the plot are useful to show, where the kinetic energy is located. The changing scales indicate the increase of energy over time. Each of the layer plots contains the 500 hPa geopotential of the unperturbed reference state. 

Perturbations structures of the shown plots for the massless kinetic energy are much smaller than for the temperature, but this is primarily an effect of the multiplication of $U$ and $V$ fields, which results in larger areas of values near zero. The normal $U$ and $V$ fields consist of similar large patterns as the temperature fields.

The PMS is also plotted and since it is not part of the variables, which we use for A-SV generation, it is initially not perturbed of course. But the variable becomes interesting, when one observes the development of the perturbations over time. This development in particular also for the unperturbed PMS shows widely meaningful structures, similar to the perturbed variables.

\begin{figure}[bt]\label{plot:RP_fc0}
\centering
\includegraphics[width=19.5cm, angle=270]{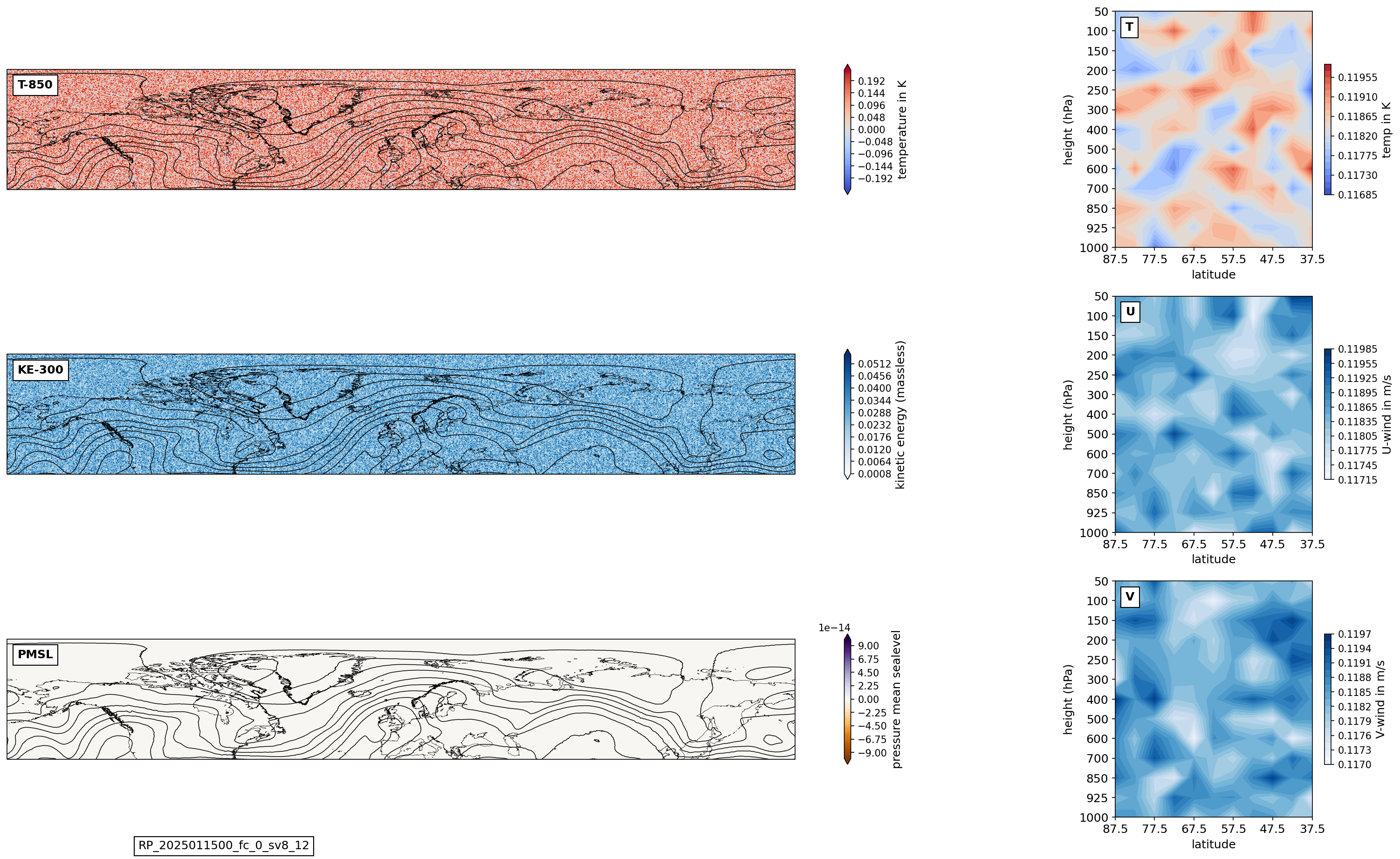}
\caption{Random perturbation, forecast leadtime 0h, increments to reference, 1) temperature 850 hPa,  2) temperature cross-section, 3) kinetic energy (massless) 300 hPa 4) U-Wind cross-section 5) pressure mean sea level (initially unperturbed) 6) V-Wind cross-section. Based on ICON $00$ UTC analysis state, $15$th January $2025$}
\end{figure}

\begin{figure}[bt]\label{plot:RP_fc24}
\centering
\includegraphics[width=19.5cm, angle=270]{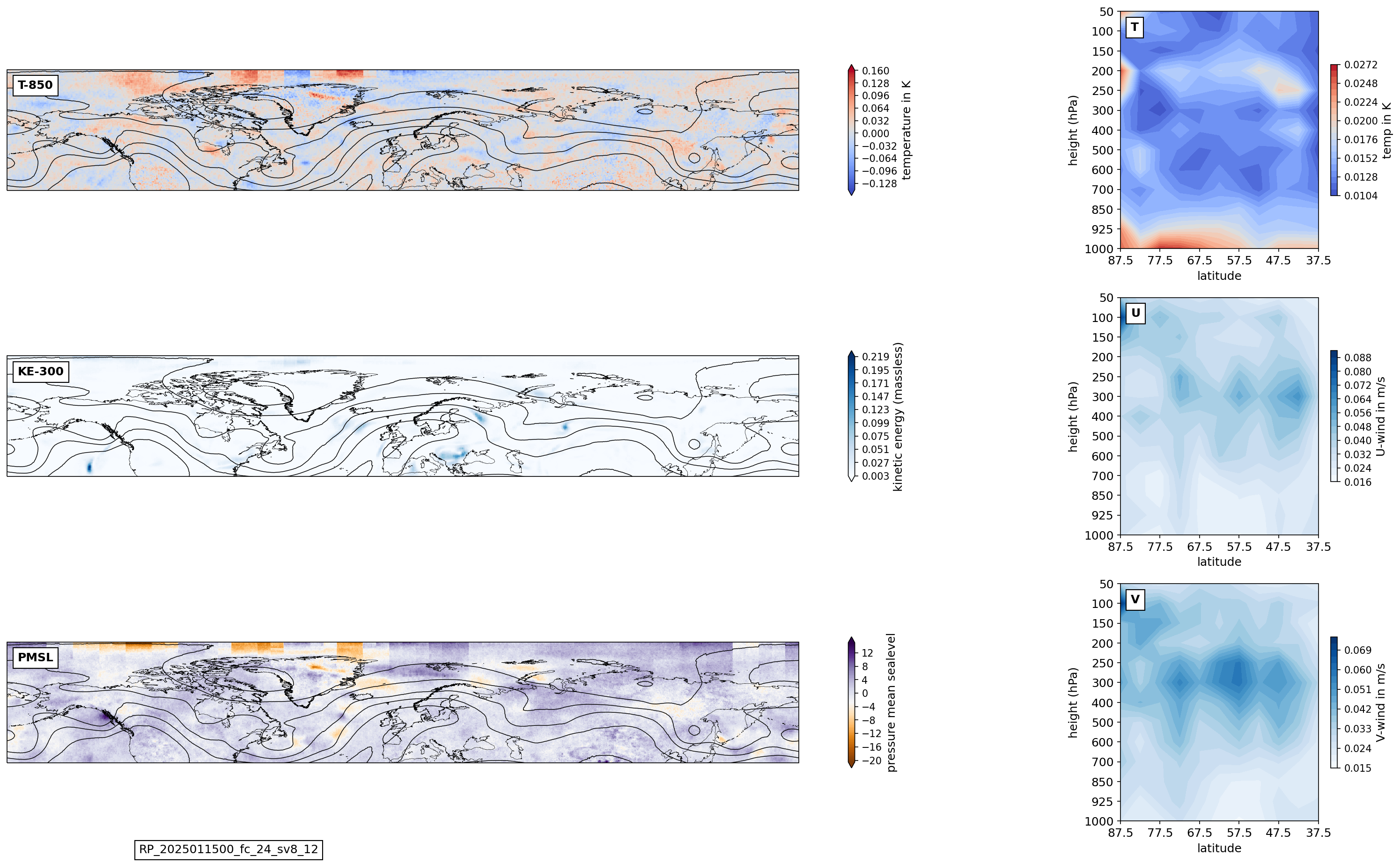}
\caption{Random perturbation, forecast leadtime 24h, increments to reference, 1) temperature 850 hPa,  2) temperature cross-section, 3) kinetic energy (massless) 300 hPa 4) U-Wind cross-section 5) pressure mean sea level (initially unperturbed) 6) V-Wind cross-section. Based on ICON $00$ UTC analysis state, $15$th January $2025$}
\end{figure}

\begin{figure}[bt]\label{plot:RP_fc48}
\centering
\includegraphics[width=19.5cm, angle=270]{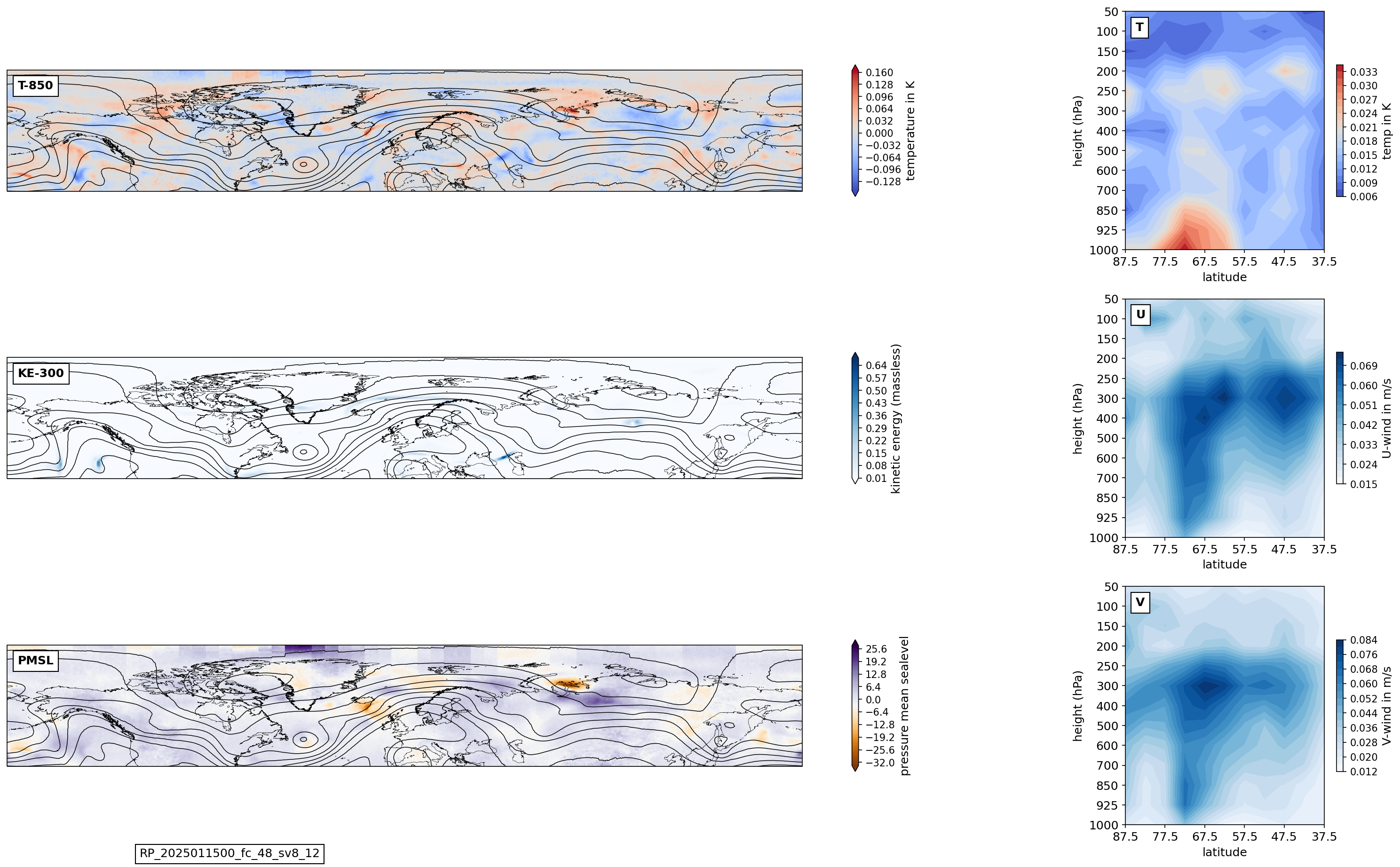}
\caption{Random perturbation, forecast leadtime 48h, increments to reference, 1) temperature 850 hPa,  2) temperature cross-section, 3) kinetic energy (massless) 300 hPa 4) U-Wind cross-section 5) pressure mean sea level (initially unperturbed) 6) V-Wind cross-section. Based on ICON $00$ UTC analysis state, $15$th January $2025$}
\end{figure}

\begin{figure}[bt]\label{plot:SV1_fc0}
\centering
\includegraphics[width=19.5cm, angle=270]{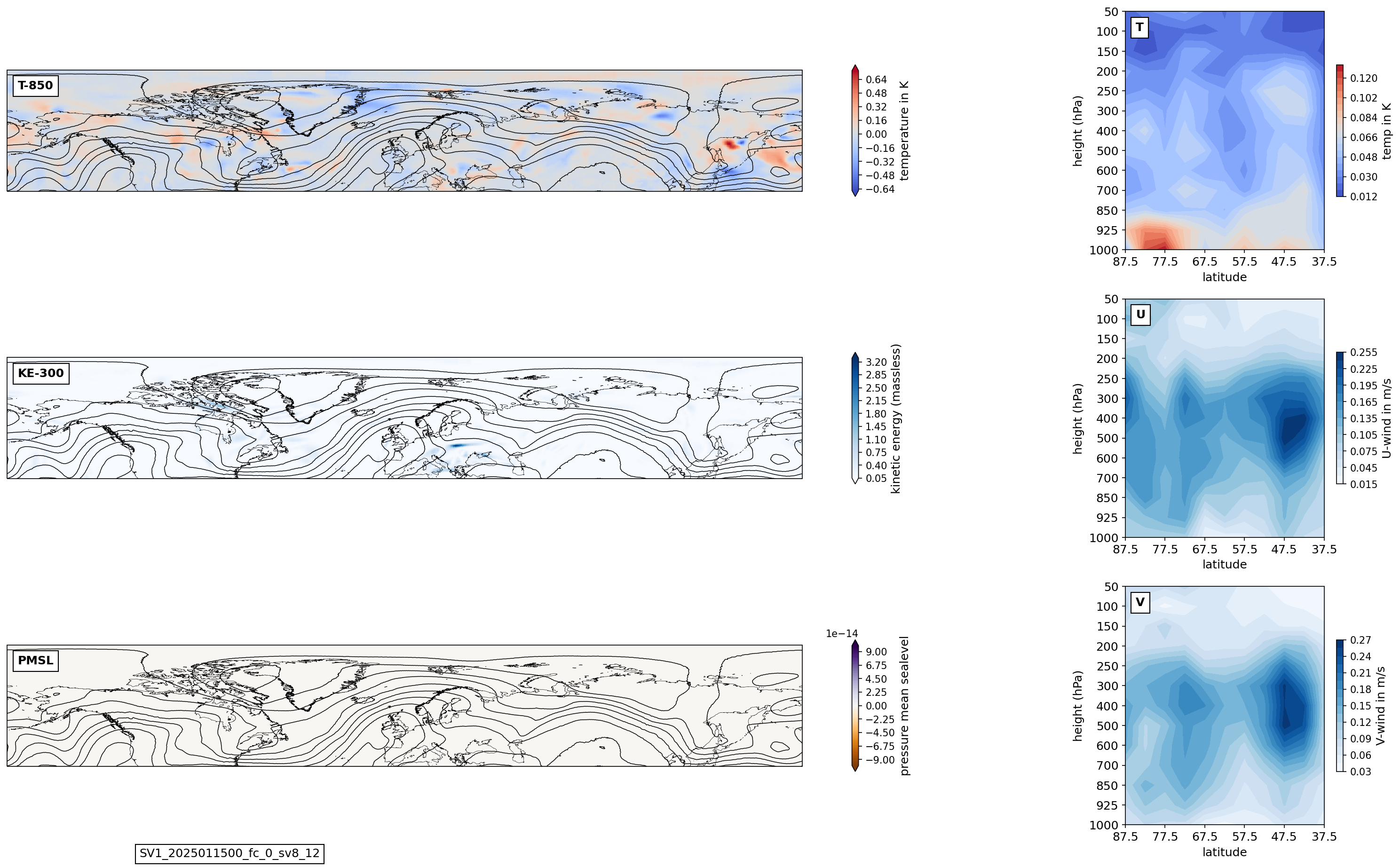}
\caption{A-SV perturbation 1.SV, forecast leadtime 0h, increments to reference, 1) temperature 850 hPa,  2) temperature cross-section, 3) kinetic energy (massless) 300 hPa 4) U-Wind cross-section 5) pressure mean sea level (initially unperturbed) 6) V-Wind cross-section. Optimization period 24h, blocksize is 8 and number of loops 12 based on ICON $00$ UTC analysis state, $15$th January $2025$}
\end{figure}

\begin{figure}[bt]\label{plot:SV1_fc24}
\centering
\includegraphics[width=19.5cm, angle=270]{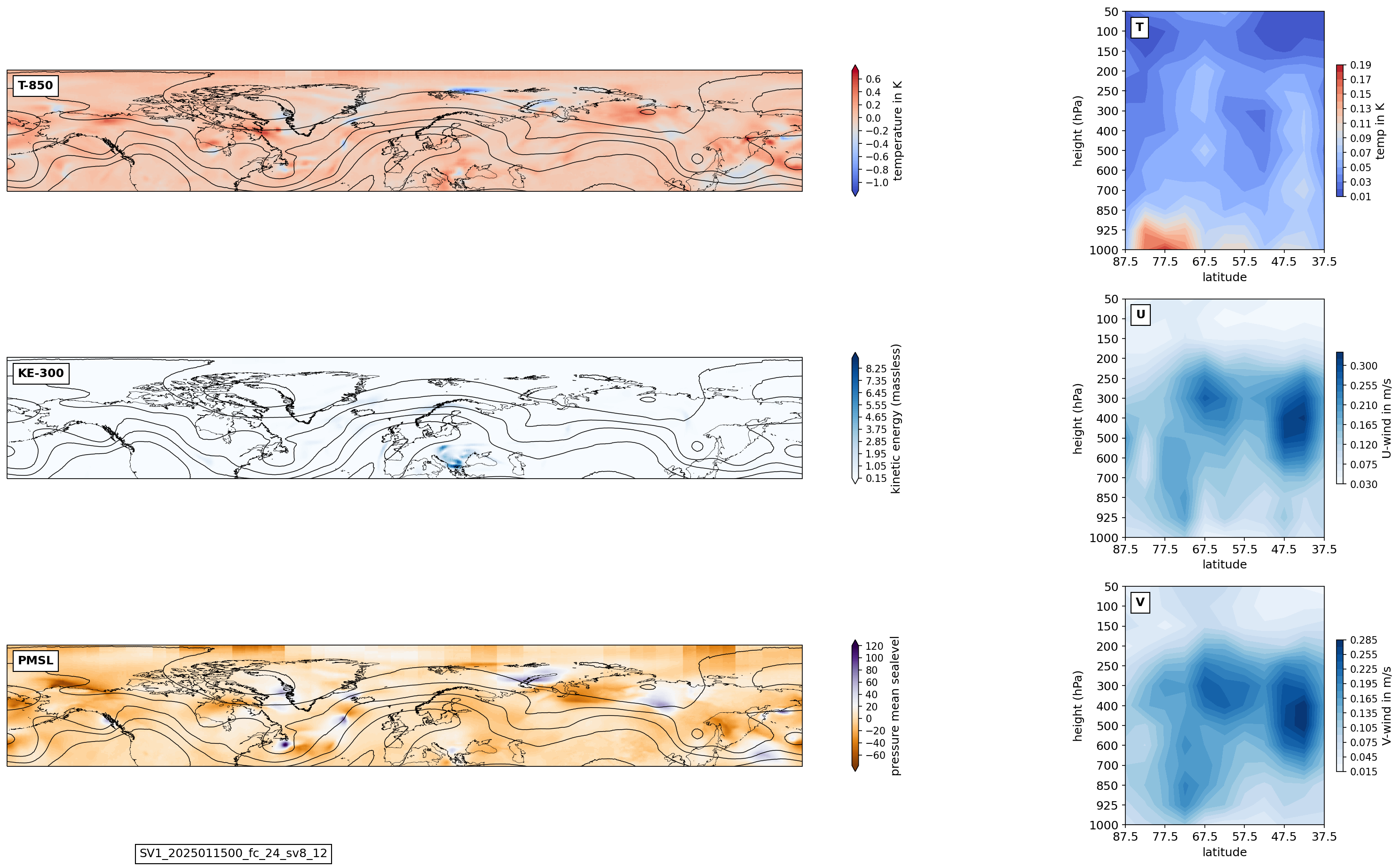}
\caption{A-SV perturbation 1.SV, forecast leadtime 24h, increments to reference, 1) temperature 850 hPa,  2) temperature cross-section, 3) kinetic energy (massless) 300 hPa 4) U-Wind cross-section 5) pressure mean sea level (initially unperturbed) 6) V-Wind cross-section. Optimization period 24h, blocksize is 8 and number of loops 12 based on ICON $00$ UTC analysis state, $15$th January $2025$}
\end{figure}

\begin{figure}[bt]\label{plot:SV1_fc48}
\centering
\includegraphics[width=19.5cm, angle=270]{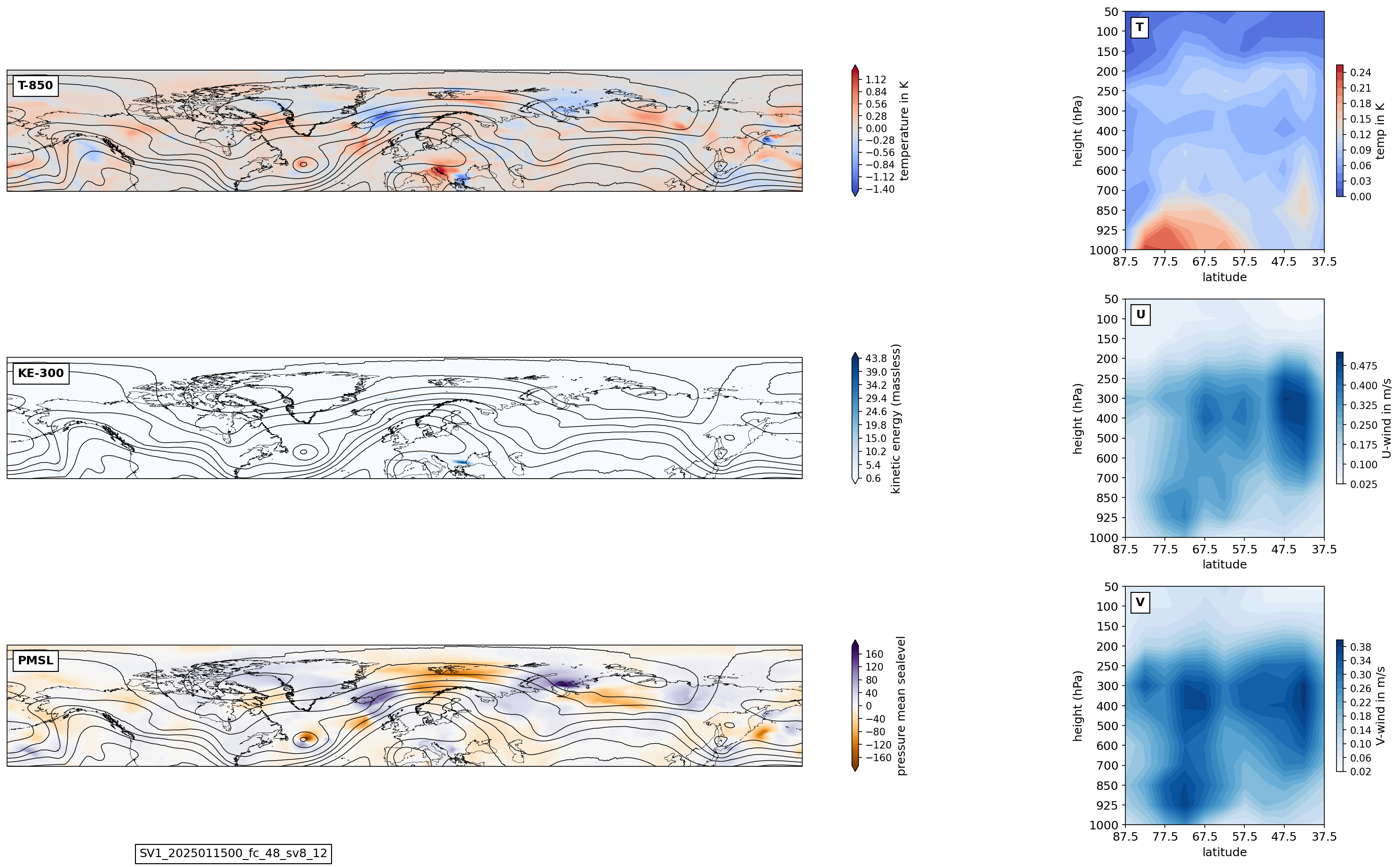}
\caption{A-SV perturbation 1.SV, forecast leadtime 48h, increments to reference, 1) temperature 850 hPa,  2) temperature cross-section, 3) kinetic energy (massless) 300 hPa 4) U-Wind cross-section 5) pressure mean sea level (initially unperturbed) 6) V-Wind cross-section. Optimization period 24h, blocksize is 8 and number of loops 12 based on ICON $00$ UTC analysis state, $15$th January $2025$}
\end{figure}

\begin{figure}[bt]\label{plot:SV1_fc72}
\centering
\includegraphics[width=19.5cm, angle=270]{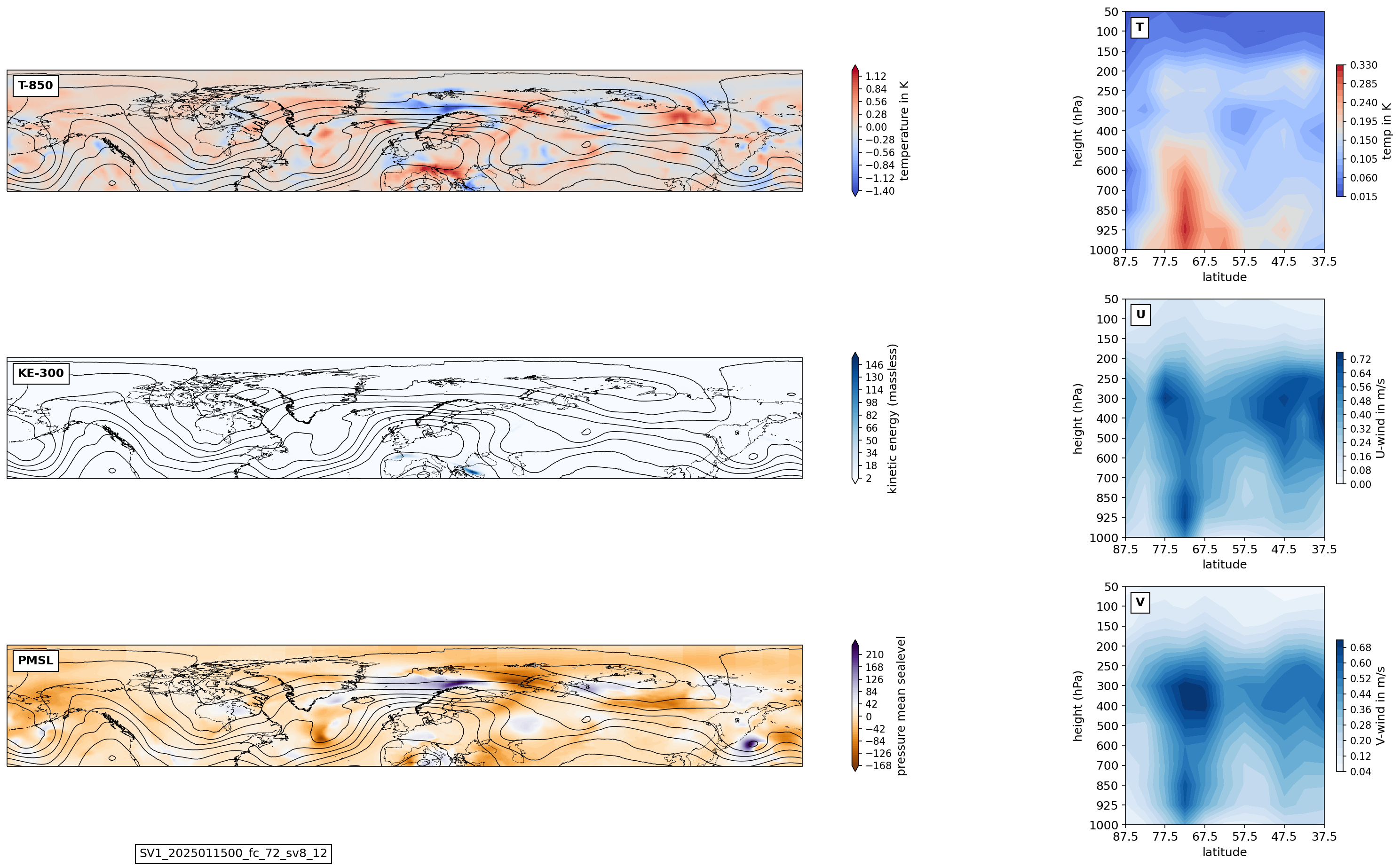}
\caption{A-SV perturbation 1.SV, forecast leadtime 72h, increments to reference, 1) temperature 850 hPa,  2) temperature cross-section, 3) kinetic energy (massless) 300 hPa 4) U-Wind cross-section 5) pressure mean sea level (initially unperturbed) 6) V-Wind cross-section. Optimization period 24h, blocksize is 8 and number of loops 12 based on ICON $00$ UTC analysis state, $15$th January $2025$}
\end{figure}

\begin{figure}[bt]\label{plot:SV1_fc96}
\centering
\includegraphics[width=19.5cm, angle=270]{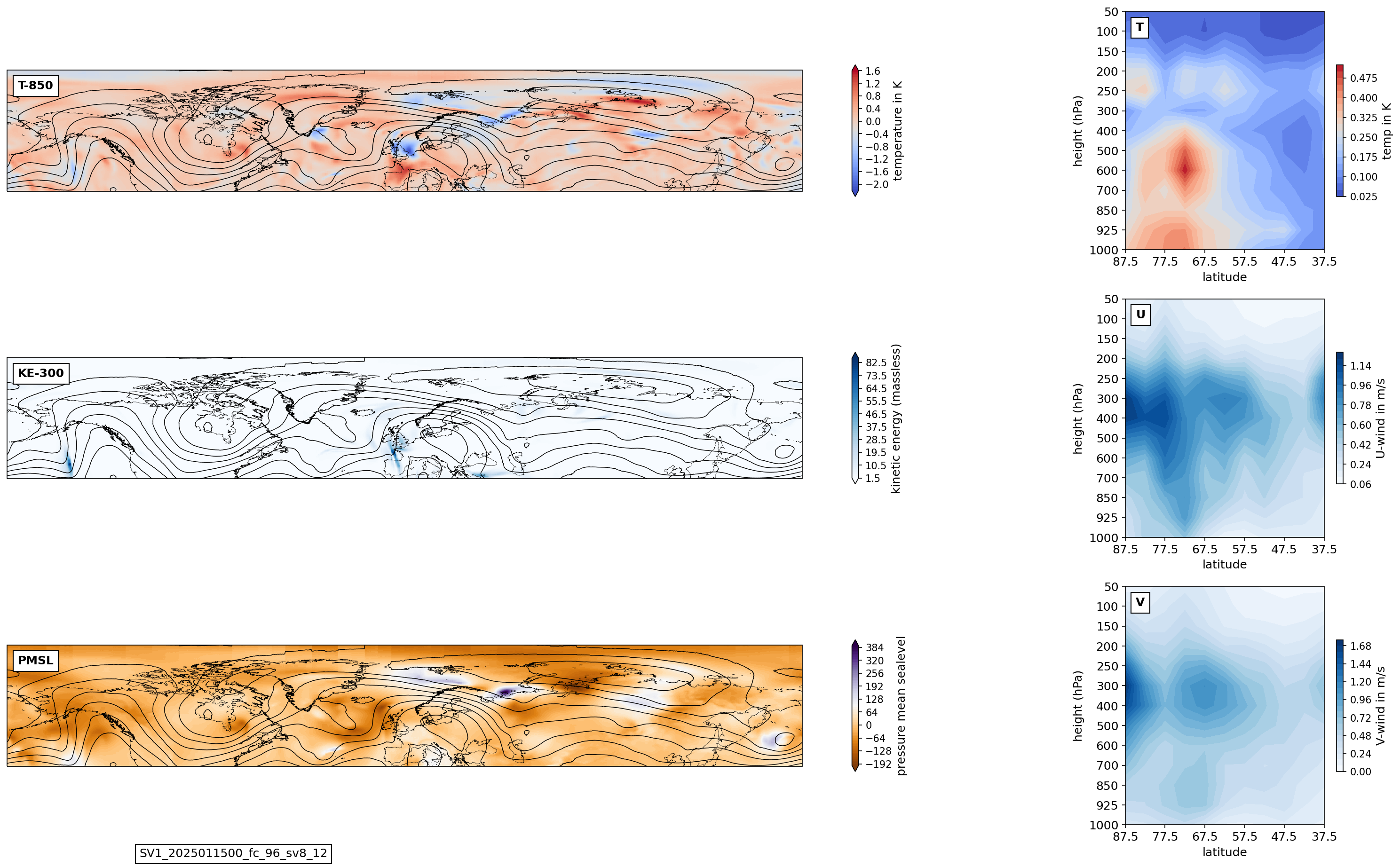}
\caption{A-SV perturbation 1.SV, forecast leadtime 96h, increments to reference, 1) temperature 850 hPa,  2) temperature cross-section, 3) kinetic energy (massless) 300 hPa 4) U-Wind cross-section 5) pressure mean sea level (initially unperturbed) 6) V-Wind cross-section. Optimization period 24h, blocksize is 8 and number of loops 12 based on ICON $00$ UTC analysis state, $15$th January $2025$}
\end{figure}

\begin{figure}[bt]\label{plot:SV1_fc120}
\centering
\includegraphics[width=19.5cm, angle=270]{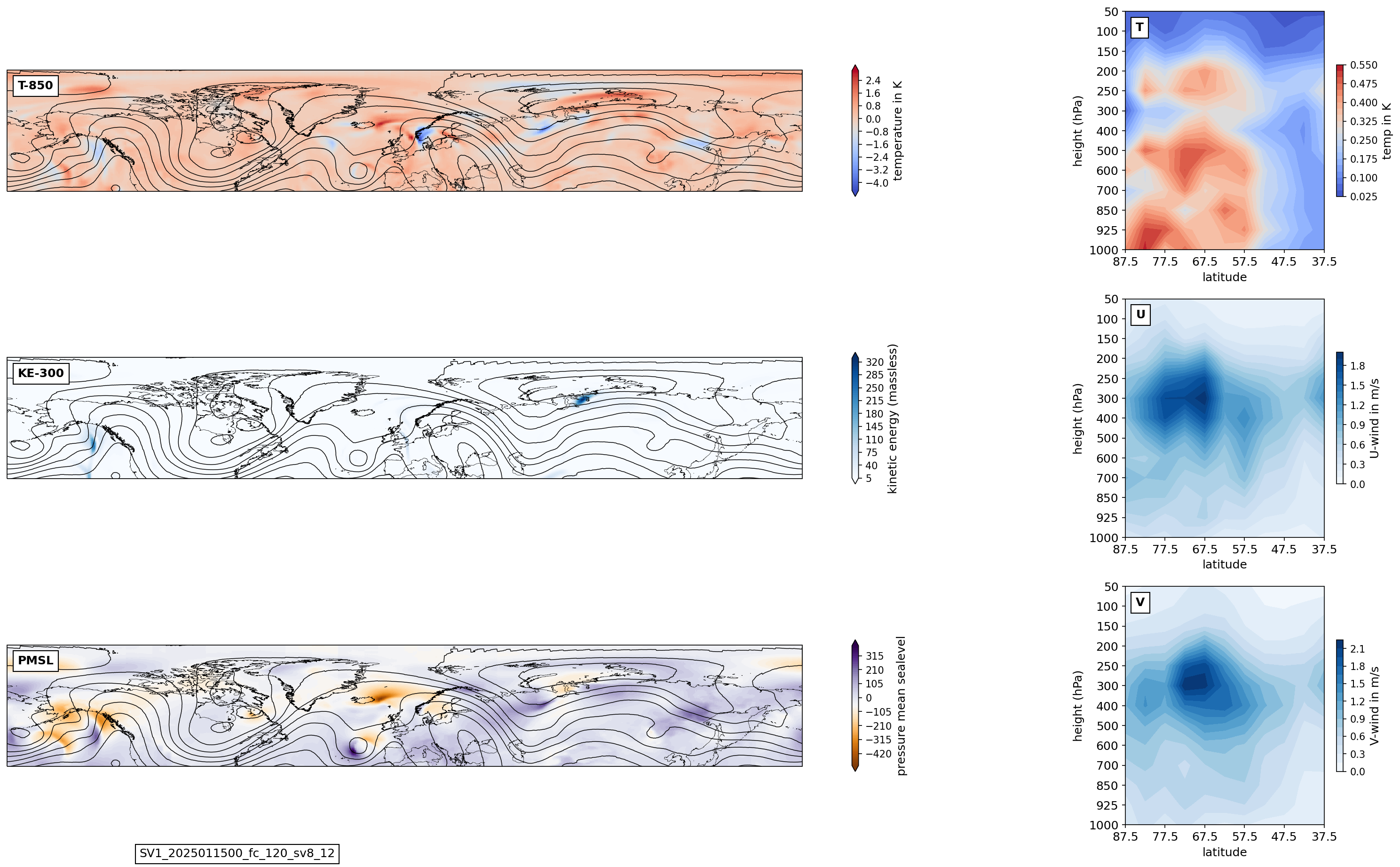}
\caption{A-SV perturbation 1.SV, forecast leadtime 120h, increments to reference, 1) temperature 850 hPa,  2) temperature cross-section, 3) kinetic energy (massless) 300 hPa 4) U-Wind cross-section 5) pressure mean sea level (initially unperturbed) 6) V-Wind cross-section. Optimization period 24h, blocksize is 8 and number of loops 12 based on ICON $00$ UTC analysis state, $15$th January $2025$}
\end{figure}

\begin{figure}[bt]\label{plot:SV1_fc144}
\centering
\includegraphics[width=19.5cm, angle=270]{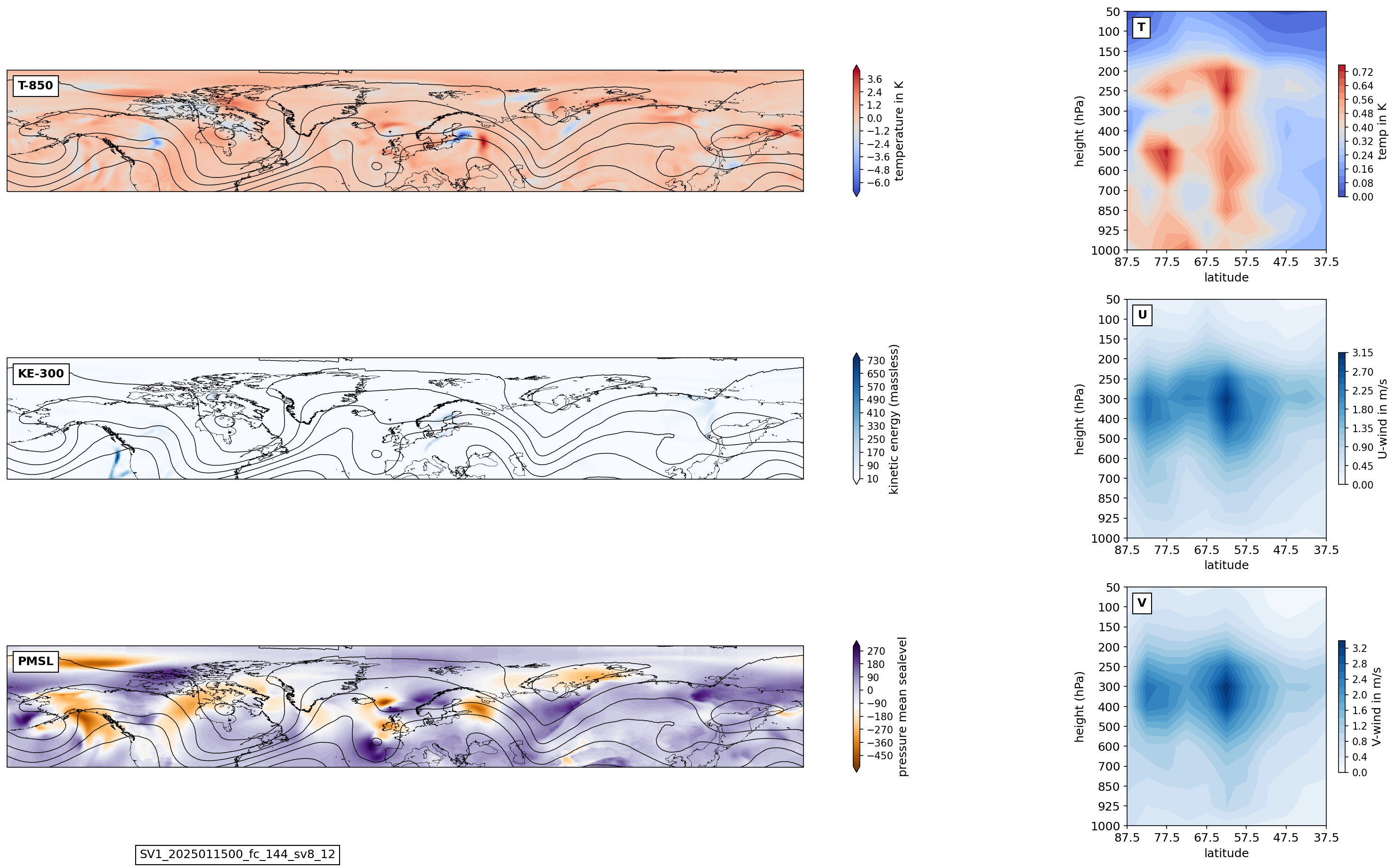}
\caption{A-SV perturbation 1.SV, forecast leadtime 144h, increments to reference, 1) temperature 850 hPa,  2) temperature cross-section, 3) kinetic energy (massless) 300 hPa 4) U-Wind cross-section 5) pressure mean sea level (initially unperturbed) 6) V-Wind cross-section. Optimization period 24h, blocksize is 8 and number of loops 12 based on ICON $00$ UTC analysis state, $15$th January $2025$}
\end{figure}

\begin{figure}[bt]\label{plot:SV1_fc168}
\centering
\includegraphics[width=19.5cm, angle=270]{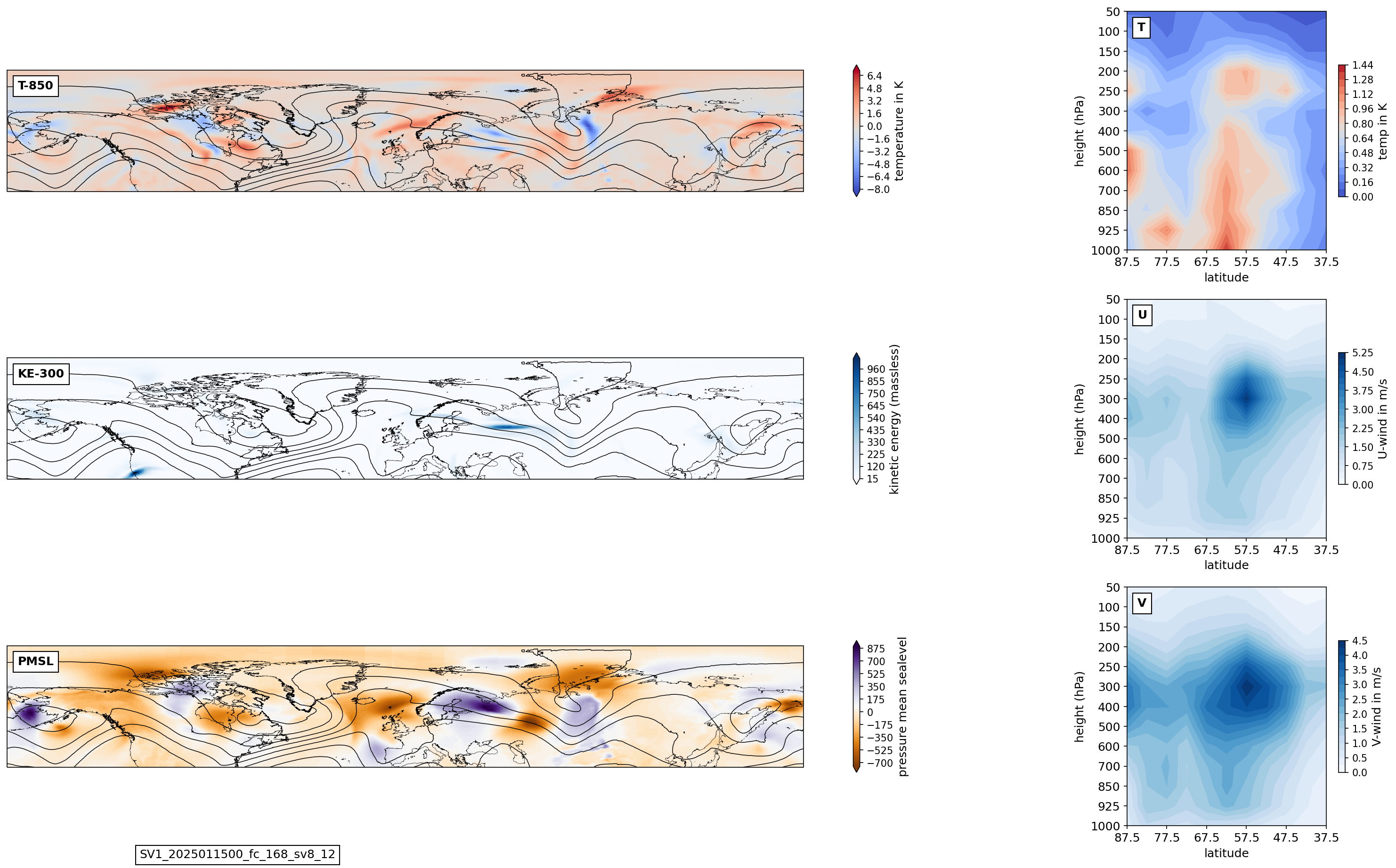}
\caption{A-SV perturbation 1.SV, forecast leadtime 168h, increments to reference, 1) temperature 850 hPa,  2) temperature cross-section, 3) kinetic energy (massless) 300 hPa 4) U-Wind cross-section 5) pressure mean sea level (initially unperturbed) 6) V-Wind cross-section. Optimization period 24h, blocksize is 8 and number of loops 12 based on ICON $00$ UTC analysis state, $15$th January $2025$}
\end{figure}

\begin{figure}[bt]\label{plot:SV17_fc0}
\centering
\includegraphics[width=19.5cm, angle=270]{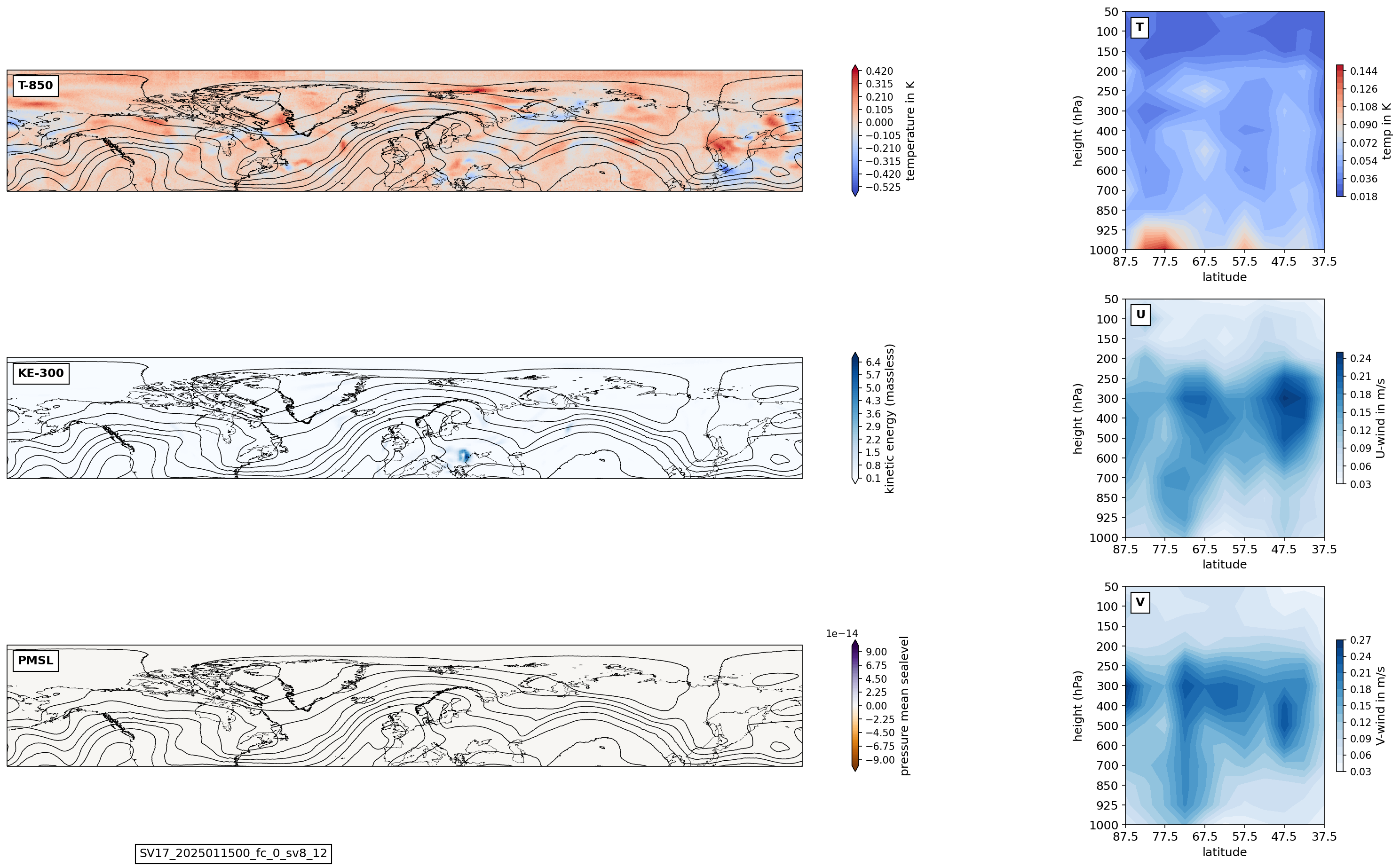}
\caption{A-SV perturbation 17.SV, forecast leadtime 0h, increments to reference, 1) temperature 850 hPa, 2) temperature cross-section, 3) kinetic energy (massless) 300 hPa 4) U-Wind cross-section 5) pressure mean sea level (initially unperturbed) 6) V-Wind cross-section. Optimization period 24h, blocksize is 8 and number of loops 12 based on ICON $00$ UTC analysis state, $15$th January $2025$}
\end{figure}

\begin{figure}[bt]\label{plot:SV17_fc24}
\centering
\includegraphics[width=19.5cm, angle=270]{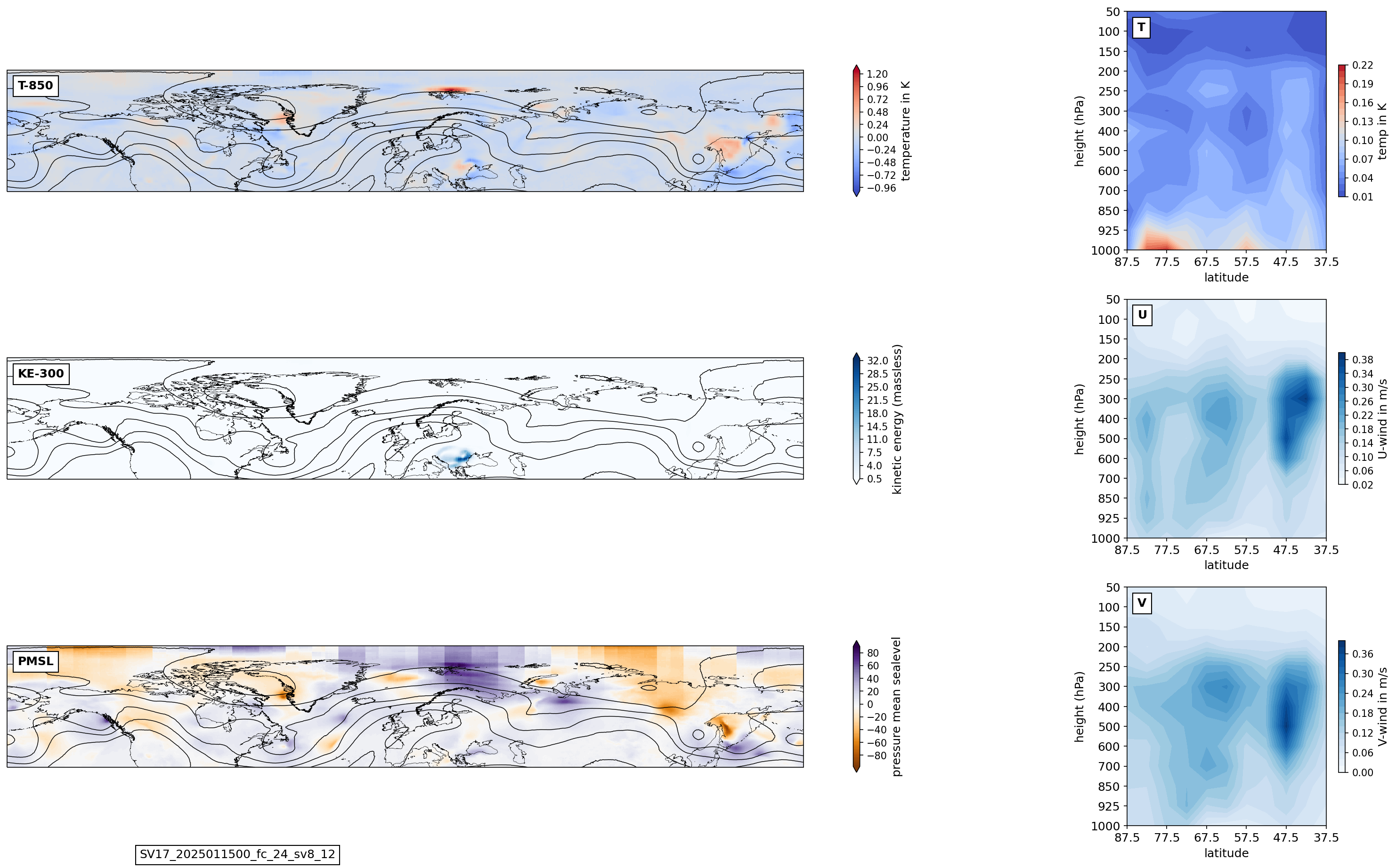}
\caption{A-SV perturbation 17.SV, forecast leadtime 24h, increments to reference, 1) temperature 850 hPa, 2) temperature cross-section, 3) kinetic energy (massless) 300 hPa 4) U-Wind cross-section 5) pressure mean sea level (initially unperturbed) 6) V-Wind cross-section. Optimization period 24h, blocksize is 8 and number of loops 12 based on ICON $00$ UTC analysis state, $15$th January $2025$}
\end{figure}

\begin{figure}[bt]\label{plot:SV17_fc48}
\centering
\includegraphics[width=19.5cm, angle=270]{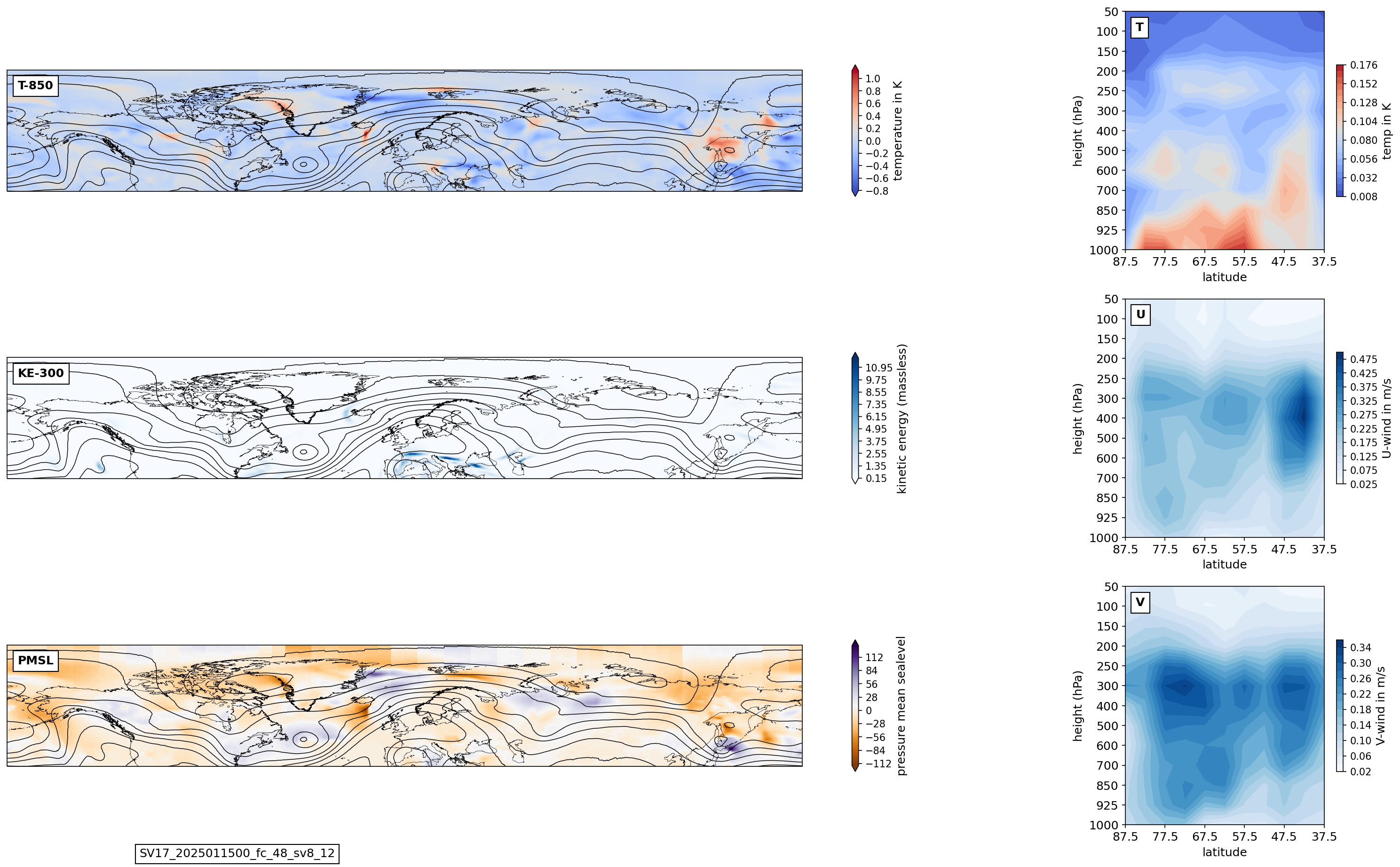} 
\caption{A-SV perturbation 17.SV, forecast leadtime 48h, increments to reference, 1) temperature 850 hPa, 2) temperature cross-section, 3) kinetic energy (massless) 300 hPa 4) U-Wind cross-section 5) pressure mean sea level (initially unperturbed) 6) V-Wind cross-section. Optimization period 24h, blocksize is 8 and number of loops 12 based on ICON $00$ UTC analysis state, $15$th January $2025$}
\end{figure}

\begin{figure}[bt]\label{plot:SV17_fc72}
\centering
\includegraphics[width=19.5cm, angle=270]{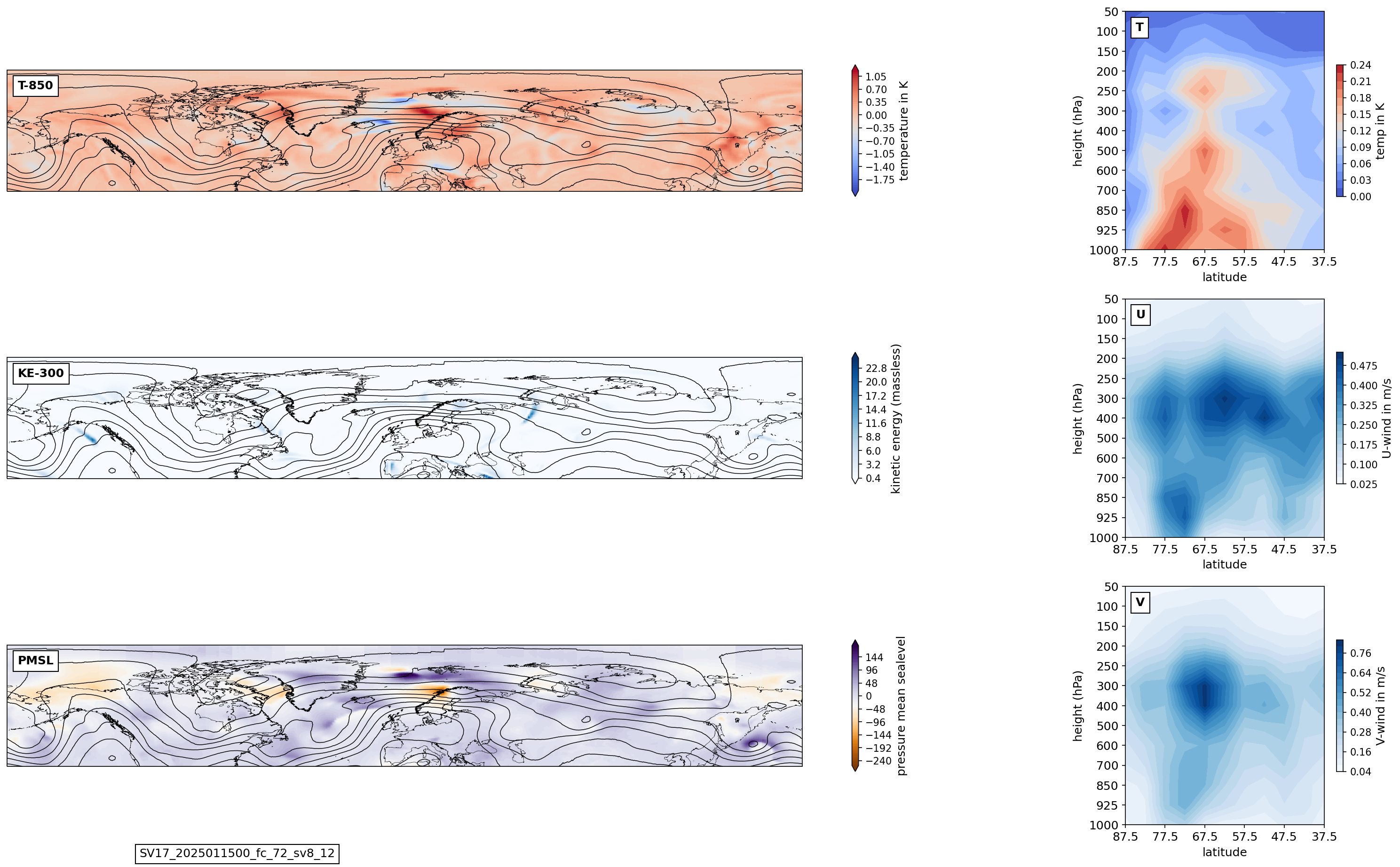}
\caption{A-SV perturbation 17.SV, forecast leadtime 72h, increments to reference, 1) temperature 850 hPa, 2) temperature cross-section, 3) kinetic energy (massless) 300 hPa 4) U-Wind cross-section 5) pressure mean sea level (initially unperturbed) 6) V-Wind cross-section. Optimization period 24h, blocksize is 8 and number of loops 12 based on ICON $00$ UTC analysis state, $15$th January $2025$}
\end{figure}

\begin{figure}[bt]\label{plot:SV17_fc96}
\centering
\includegraphics[width=19.5cm, angle=270]{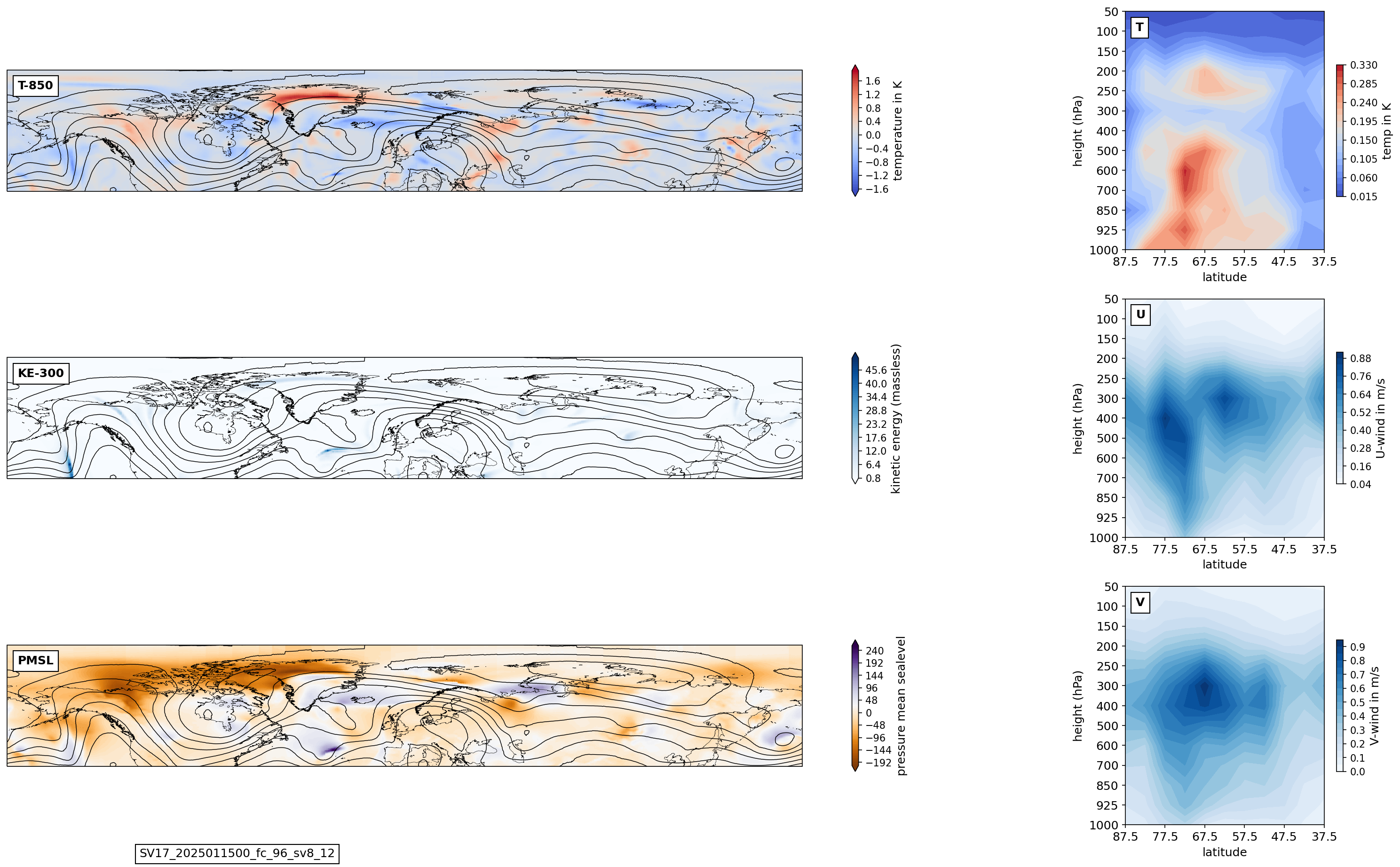}
\caption{A-SV perturbation 17.SV, forecast leadtime 96h, increments to reference, 1) temperature 850 hPa, 2) temperature cross-section, 3) kinetic energy (massless) 300 hPa 4) U-Wind cross-section 5) pressure mean sea level (initially unperturbed) 6) V-Wind cross-section. Optimization period 24h, blocksize is 8 and number of loops 12 based on ICON $00$ UTC analysis state, $15$th January $2025$}
\end{figure}

\begin{figure}[bt]\label{plot:SV17_fc120}
\centering
\includegraphics[width=19.5cm, angle=270]{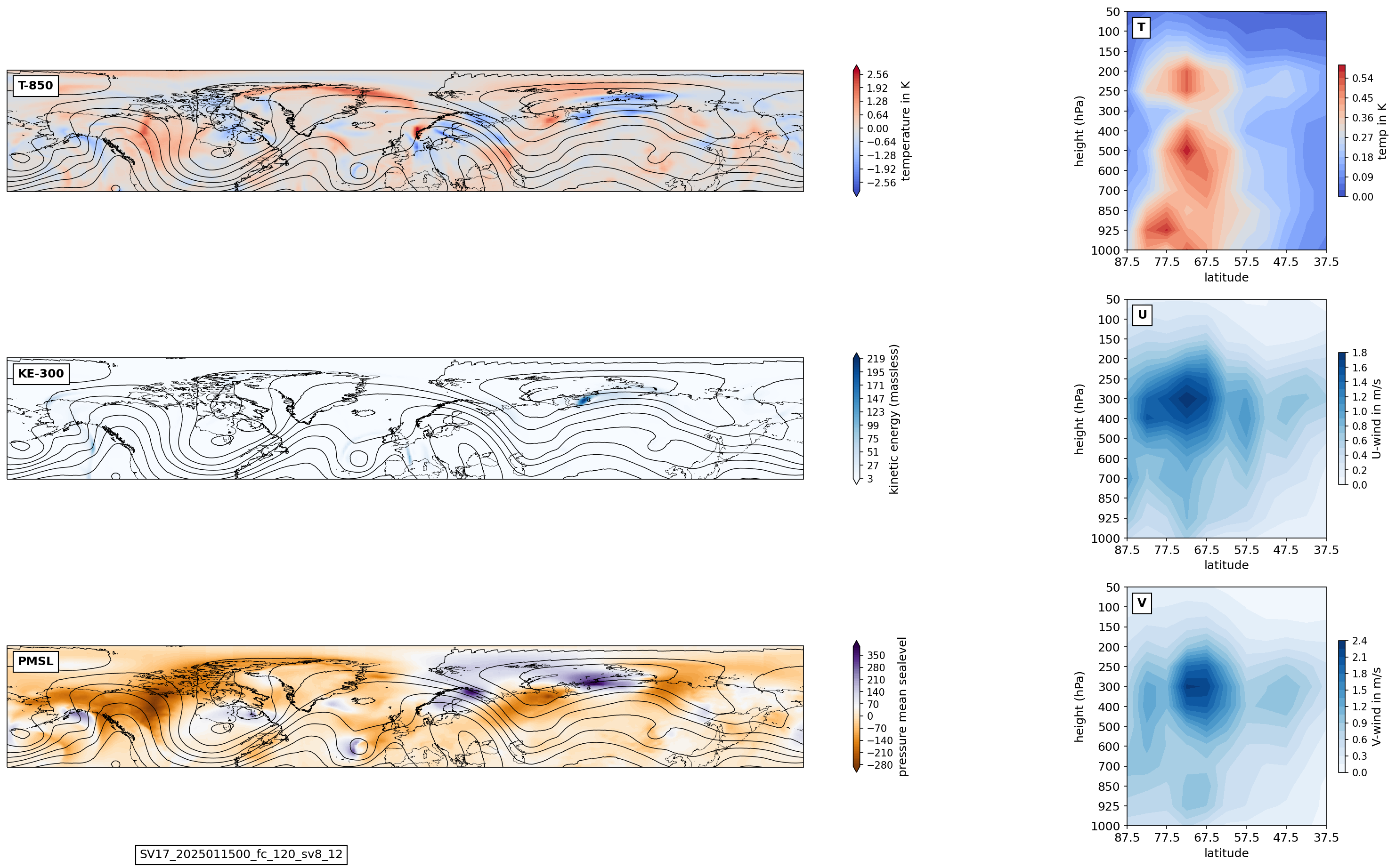}
\caption{A-SV perturbation 17.SV, forecast leadtime 120h, increments to reference, 1) temperature 850 hPa, 2) temperature cross-section, 3) kinetic energy (massless) 300 hPa 4) U-Wind cross-section 5) pressure mean sea level (initially unperturbed) 6) V-Wind cross-section. Optimization period 24h, blocksize is 8 and number of loops 12 based on ICON $00$ UTC analysis state, $15$th January $2025$}
\end{figure}

\begin{figure}[bt]\label{plot:SV17_fc144}
\centering
\includegraphics[width=19.5cm, angle=270]{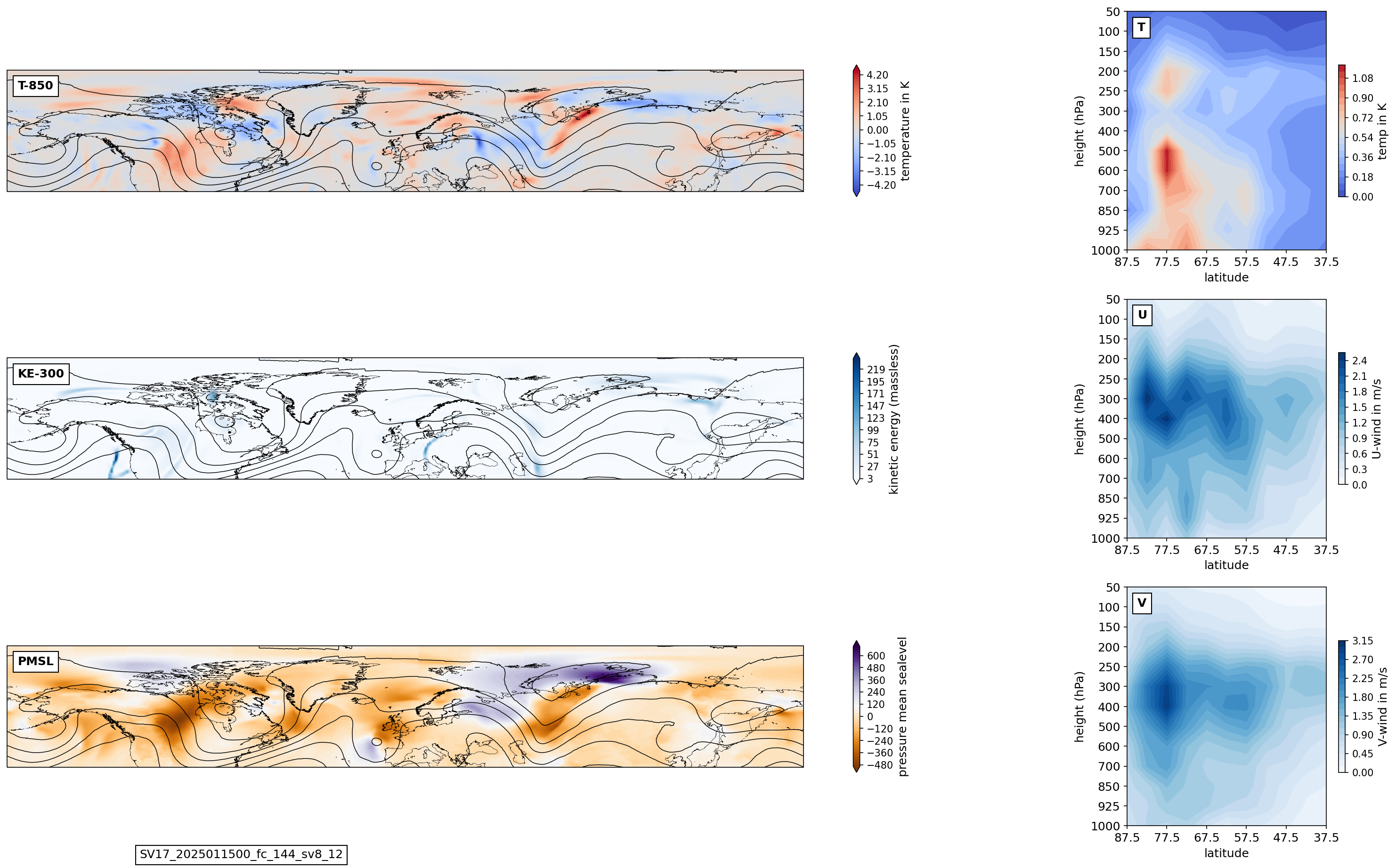}
\caption{A-SV perturbation 17.SV, forecast leadtime 144h, increments to reference, 1) temperature 850 hPa, 2) temperature cross-section, 3) kinetic energy (massless) 300 hPa 4) U-Wind cross-section 5) pressure mean sea level (initially unperturbed) 6) V-Wind cross-section. Optimization period 24h, blocksize is 8 and number of loops 12 based on ICON $00$ UTC analysis state, $15$th January $2025$}
\end{figure}

\begin{figure}[bt]\label{plot:SV17_fc168}
\centering
\includegraphics[width=19.5cm, angle=270]{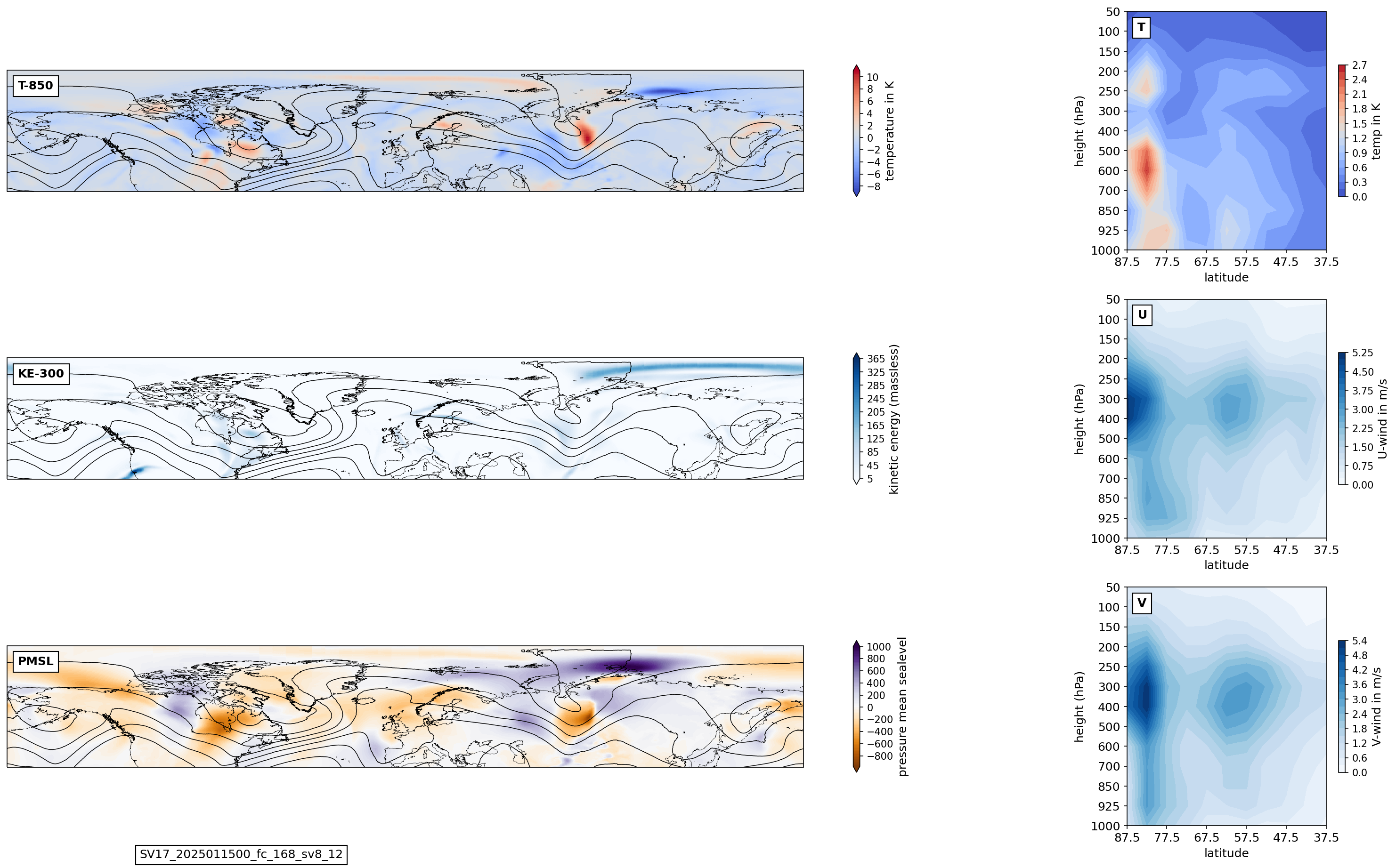}
\caption{A-SV perturbation 17.SV, forecast leadtime 168h, increments to reference, 1) temperature 850 hPa, 2) temperature cross-section, 3) kinetic energy (massless) 300 hPa 4) U-Wind cross-section 5) pressure mean sea level (initially unperturbed) 6) V-Wind cross-section. Optimization period 24h, blocksize is 8 and number of loops 12 based on ICON $00$ UTC analysis state, $15$th January $2025$}
\end{figure}

\end{document}